\documentstyle[epsf,graphicx,11pt]{article}

\newcommand{\Ref}[1]{(\ref{#1})}

\newcommand{\cmod}[1] {|{#1}|^2}
\newcommand{\acc}{\\[3mm]}

\newcommand\ph{\varphi_{i}}
\newcommand\phl{\varphi_{i-1}}
\newcommand\phr{\varphi_{i+1}}

\newcommand\uc{u_{i}}
\newcommand\ul{u_{i-1}}
\newcommand\ur{u_{i+1}}
\newcommand\ull{u_{i-2}}
\newcommand\urr{u_{i+2}}

\newcommand\ssc{s_{i}}
\newcommand\ssl{s_{i-1}}
\newcommand\ssr{s_{i+1}}
\newcommand\ssll{s_{i-2}}
\newcommand\ssrr{s_{i+2}}

\begin{document}

\title{Electron self-trapping on a nano-circle\\
}
\author{
L.S. Brizhik\thanks{e-mail address: brizhik@bitp.kiev.ua},\,
and
A.A. Eremko\thanks{e-mail address: eremko@bitp.kiev.ua},\,
\\
Bogolyubov Institute for Theoretical Physics, 03143 Kyiv, Ukraine
\acc
\acc
B. Piette\thanks{e-mail address: B.M.A.G.Piette@durham.ac.uk} \,
and
W. Zakrzewski\thanks{e-mail address: W.J.Zakrzewski@durham.ac.uk}
\\
Department of Mathematical Sciences, University of Durham, \\
Durham DH1 3LE, UK\\
}
\date{}
\maketitle

\begin{abstract}
We study the self-trapping of quasiparticles (electrons, holes, excitons, etc) 
in a molecular chain with the structure of a ring, taking into account
the electron-phonon interaction and the radial and tangential deformations 
of the chain.
A discrete system of equations is obtained and solved numerically.
The analytical solutions for the wave function of a quasiparticle 
 and for the molecule displacements that determine the
distortion of the ring, are also obtained and solved in the continuum 
approximation.
The numerical solutions of the system of discrete nonlinear equations reveals
several regimes of quasiparticle localisation in the chain which depend on
the values of the parameters of the system. It is shown that the transversal 
deformation of the chain favours the formation of a soliton.

\end{abstract}
\section{Introduction.}


The study of various physical properties of carbon nanotubes
has received a lot of attention during the last few years.
In particular, B. Hartmann et al \cite{BH_WJZ} have recently
studied the electron-phonon interaction on a hexagonal lattice.
Their approach was, however, limited in its scope as it 
did not take into account the
possible displacement of atoms in the direction perpendicular 
to the nanotube. To address this problem, we have to introduce three  
phonon fields, two of which are tangent to the tube while the third 
one is perpendicular to it. The introduction of such a transversal 
field is not very straightforward. It 
requires the 
generalization of the phonon Hamiltonian to describe the bending 
elasticity of the tube. Such a generalization is 
presented in this paper for the case of a nanocircle. A study of the 
electron-phonon interaction in a simple model where such a 
transversal field is introduced, paves the way towards the study of 
a more relevant model describing carbon nanotubes. 
It also describes various  physical  
properties, including the conductivity and conformational states, 
of other physical systems such as 
circular macromolecules, cyclic organic, inorganic and semi-inorganic 
polymeric macromolecules. 

 Our model 
describes a circular closed chain of monomers which are allowed to move 
in the plane spanned by the circle at rest.  Thus, it is a 
two-dimensional model where deformations in the 
direction perpendicular to the circle plane, are not allowed.  
One can think of this 
model as a crude cross section of a carbon nanotube, though 
obviously, the hexagonal symmetry of the carbon nanotube, and its 
three-dimensional nature require a more general description.
Nevertheless, our hope is to gain some understanding of 
how one can take into account the transversal displacement of atoms.

At the same time, our model does describe 
real physical low-dimensional cyclic nanosystems that 
include a large class of end-linked natural 
and synthetic low-dimensional macromolecules and nanosystems with a 
circular structure. Amongst them
one finds the cyclic biological macromolecules, such as intact DNA from 
most prokaryotes and viruses, mitochondria plasmid DNA 
\cite{stryer,swenb,benham}, cyclic DNA macroarrays used for the 
systematic screening of gene expression \cite{freeman,blesz,seeman}.  
Large circular single stranded RNAs form genomic material of viroids 
\cite{zaug,sech}. Some of the oligoribonucleotides with 
known biological activity are also cyclic molecules 
\cite{ross,hsu,allaz,fried}.   
The so called ``second messengers" also have a circular structure.  
Amongst them are the messengers which regulate the channels involved in the 
initiation of 
impulses in neurons responding to odors and light, messengers which 
function as calcium regulated transcription factors or as a substrate 
for depolarization-activated calcium calmodulin-dependent protein 
kinases I and II (e.g., cAMP Response Element-Binding Protein, known 
also as CREB protein) \cite{creb}. In addition, there are
also  messengers which induce and 
control transcription from the proximal promoter of the human 
aromatase responsible for the metabolism of steroids 
\cite{michael}. Therefore, our study of distortions of molecular 
nano-circles due to electron-phonon coupling could bring some insight 
into the structure-function relationship of biological 
macromolecules.


End-linked organic, inorganic and semi-inorganic 
polymers are now known  as functional materials 
with applications as biologically active species, catalyst 
supports, charge carriers etc.  For instance, the long-chain 
circular esters are used for soft lenses, cyclic trimeric phosphazene 
rings are valuable technological materials for their high heat 
resistance and flame retardant behaviour.  Polysiloxanes, which are 
the main components of silicone products, are one of the most 
important cyclic materials (\cite{mark} and references cited 
therein).


The circular structure of these macromolecules, 
their relative softness, deformability and ability to form 
cross-linking suggest modelling such systems by a nanocircle of 
periodically placed monomers of a given length. Under certain 
conditions one or several monomers of the nanocircle can be excited by 
an external field (light or electromagnetic radiation, trapping of 
extra electron or hole, etc.). Such an excitation, called in what follows, a
quasiparticle, can move onto the neighbouring monomers and cause a local 
distortion of the ring due to the 
electron-phonon coupling. In a linear chain it can lead to the 
change of properties of the quasiparticle, depending on the 
strength of the coupling. At weak coupling the quasiparticle remains 
almost free, at strong coupling it is trapped on one site (a small 
polaron is formed), while at intermediate values of the coupling 
a self-trapping 
occurs within few lattice sites and a nonlinear soliton-like state is 
formed \cite{ac+op,superc,synmet3}. It turns out that the  
physical properties of systems at different regimes of localisation differ 
significantly, and, therefore, the classification of systems by their 
electron ground states is important.

One can expect that, depending on the strength of the electron-phonon
coupling, similar regimes take place in a very large
end-linked circle. However, this is not clear for a circle
of a finite and not very large length. This problem is studied in the present
paper. We start by deriving a Hamiltonian describing the elastic properties of
a nanocircle and a Hamiltonian describing the interaction between 
the elecron and the phonon fields. We then derive the corresponding  
system of discrete equations and compute solutions for this model. 
We show that the model exhibits several classes of solutions 
depending on the values of the parameters. In particular, we show 
that the electron-phonon interaction leads to the self-trapping of a 
quasiparticle and introduces a deformation of the chain.

\section{The nano-circle}

We consider a molecular chain which consists of periodically repeated
units (atoms or groups of atoms) bound by chemical or hydrogen bonds. We
assume that these bonds between the units are directed in such a way
that, for a molecular chain with free ends, the minimum of the free
energy corresponds to a linear one-dimensional structure in which
chain sites are placed at the equilibrium distance
$a_0$ and all intersite bonds are directed along the chain axis.

We focus our attention on a study of the molecular chain which
is closed to form a circle (before any deformations are allowed).
The equilibrium positions of the chain sites are
determined by the radius-vectors
\begin{equation}
\vec{R}_n^{(0)} = R_0 [\vec{e}_x \sin(n\alpha + \theta_0) + \vec{e}_y
\cos(n \alpha + \theta _0) ]
\end{equation}
where the centre of the coordinate system is placed at
 the centre of the circle,
$R_0$ is the circle radius, $n = 1,\dots ,N$ enumerates the chain sites,
$\alpha = 2\pi/N$ where $N$ is the number of sites in the ring,
and $\theta _0$ is an arbitrary angle.
The equilibrium distance between the sites in the molecular circle is
\begin{equation}
D^{(0)}_{n,n+1} = |\vec{D}^{(0)}_{n,n+1}| = |\vec{R}^{(0)}_{n+1}
-\vec{R}^{(0)}_{n}| = 2 R_0 \sin(\frac{\alpha}{2}) = a
\end{equation}
which, in general, can differ from $a_0$.
Thus, the radius of the circle is $R_0 = a/(2\sin(\frac{\alpha}{2}))$.
Note that if there is a bending of chemical bonds at each site $n$, this
bending can be described by the angle $\beta^{(0)}_n$ between the
vectors $\vec{D}^{(0)}_{n-1,n} = \vec{R}^{(0)}_{n} -
\vec{R}^{(0)}_{n-1}$ and $\vec{D}^{(0)}_{n,n+1} = \vec{R}^{(0)}_{n+1}
- \vec{R}^{(0)}_n$:
\begin{equation}
\cos\beta^{(0)}_n =
{ {(\vec{D}^{(0)}_{n-1,n} \cdot \vec{D}^{(0)}_{n,n+1})}\over
{|\vec{D}^{(0)}_{n-1,n}| |\vec{D}^{(0)}_{n,n+1}| }} = \cos(\alpha),
\end{equation}
i.e., $\beta^{(0)}_n = \alpha$.

Due to the displacements the real positions of the sites are
\begin{equation}
\vec{R}_n = \vec{R}_n^{(0)} + \vec{r}_n,
\end{equation}
where $\vec{r}_n$ is the displacement of the site $n$ from the equilibrium
position. Note that we can represent the vector $\vec{r}_n$ using the
following
orthogonal unit vectors $\vec{e}_{n}^{(r)}$ and $\vec{e}_{n}^{(t)}$
\begin{eqnarray}
\vec{e}_{n}^{(t)} &=& \vec{e}_x \cos(n\alpha + \theta_0) - \vec{e}_y
\sin(n \alpha + \theta_0), \\
\vec{e}_{n}^{(r)} &=& \vec{e}_x \sin(n\alpha + \theta_0) + \vec{e}_y
\cos(n \alpha + \theta_0).
\end{eqnarray}
We then have
\begin{equation}
\vec{r}_{n} = \vec{e}_{n}^{(t)} r_{t,n} + \vec{e}_{n}^{(r)}r_{r.n},
\end{equation}
where  $r_{t,n} \equiv u_n$ is the displacement tangential to the circle
while $r_{r,n}\equiv s_n$ is
the displacement perpendicular to the circle (the tangential and
radial components of displacement respectively). Thus the real
distances between the sites and the bending angles differ from
their equilibrium values: $D_{n,n+1} = a + d_{n,n+1}$ and
$\beta_n = \alpha + \Delta \beta _n $.

The vector, which links the site $n$ with the $n+1$ one, is
$ \vec{D}_{n,n+1} = \vec{R}_{n+1} - \vec{R}_n =
\vec{D}^{(0)}_{n,n+1} + \vec{d}_{n,n+1}$ where $\vec{d}_{n,n+1} =
\vec{r}_{n+1} - \vec{r}_{n}$. So, the distance is given by
\begin{equation}
D_{n,n+1} = |\vec{D}_{n,n+1}| = \sqrt{(\vec{D}^{(0)}_{n,n+1} +
\vec{d}_{n,n+1})^2} \approx a + {{(\vec{D}^{(0)}_{n,n+1} \cdot
\vec{d}_{n,n+1})}\over{a}} = a + U_n,
\end{equation}
where
\begin{equation}
U_n = \cos(\frac{\alpha}{2}) (u_{n+1}-u_n) +
       \sin(\frac{\alpha}{2}) (s_n+ s_{n+1}).
\label{Ui}
\end{equation}

To obtain this expression we have assumed that the relative site
displacements are small, i.e. ${\cmod{\vec{d}_{i,i+1}}} \ll a^2$.
In this case the deviations of the bending angles from their equilibrium
values are also small, and we can write
\begin{eqnarray}
\cos(\beta_n) &=& \cos( \alpha + \Delta \beta_n) \\
&=& \cos(\alpha) \cos(\Delta \beta_n) - \sin(\alpha) \sin(\Delta
 \beta_n) \approx \cos(\alpha) - \sin(\alpha) (\Delta \beta_n). \nonumber
\end{eqnarray}
For $\cos(\beta_n)$ we can put
\begin{eqnarray}
\cos\beta_n & = &{(\vec{D}_{n-1,n} \cdot \vec{D}_{n,n+1})}\over
{|\vec{D}_{n-1,n}| |\vec{D}_{n,n+1}| } \nonumber\\
&\approx & {(\vec{D}^{(0)}_{n-1,n} \cdot \vec{D}^{(0)}_{n,n+1}) +
(\vec{D}^{(0)}_{n-1,n} \cdot \vec{d}_{n,n+1}) +
(\vec{D}^{(0)}_{n,n+1} \cdot \vec{d}_{n-1,n})}\over
{(a + U_{n-1})\,(a + U_n)}
\\
&\approx &\cos\alpha - \sin\alpha \left(\sin\frac{\alpha}{2}
\frac{u_{n+1}-u_{n-1}}{a} + \cos\frac{\alpha}{2} \frac{2s_n -
s_{n+1} - s_{n-1}}{a} \right).\nonumber
\end{eqnarray}

Thus, we have arrived at the following expression for $\Delta \beta_n$:
\begin{equation}
\Delta \beta_n = \frac{A_n}{a},
\end{equation}
where
\begin{equation}
A_n = \sin\frac{\alpha}{2} (u_{n+1}-u_{n-1}) + \cos\frac{\alpha}{2}
(2s_n - s_{n+1} - s_{n-1}).
\label{Ai}
\end{equation}

\section{The Hamiltonian}

Next, we assume that the individual isolated monomers are characterised
by nondegenerate electronic excitations with energy $E_0$.
In the approximation of the nearest neighbour interaction, the total
Hamiltonian which describes the electronic excitations (or, in the tight 
binding approximation, the additional electrons) and harmonic vibrations 
in a molecular chain can be written as
\begin{equation}
 H = \sum_{n} \Bigl(
{\cal E}_n B^{\dag}_n B_n
-\,J_{n,n+1}\,(B^{\dag}_n B_{n+1} + B^{\dag}_{n+1} B_n )\Bigr) + H_{ph},
\end{equation}
where $\ B^{\dag}_n,\  B_n$ are the creation and annihilation
operators of the quasiparticle at the $n$-th site of the circle,
${\cal E}_n = E_0 + V_n$ is the on-site electron energy, 
where $V_n$ is the change of the electron energy due
to the influence of the neighbouring sites. Furthermore,  $J_{n,n+1}$
describes the quasiparticle transfer 
between the nearest neighbours, and $H_{ph}$ is the Hamiltonian of the site
displacements. The electron-phonon interaction is determined by the
dependence of $V_n$ and $J_{n,n+1}$ on the positions of the neighbours of
the $n$-th site. The term $J_{n,n+1}$ depends on the
distance between the sites $D_{n,n+1} = a + U_n$ and, within the linear
approximation with respect to the displacements, we can put it as
\begin{equation}
J_{n,n+1} = J - G_2 U_n, 
\end{equation}
where $J = J(a)$ and $G_2 = - dJ/da$.

$V_n$ is determined by the influence of the neighbours; thus it depends on
the distance between them and on the bending angle of the chemical bonds:
$V_n = V_{n-1,n}(a + U_{n-1}) + V_{n,n+1}(a+U_n)+V_n(\beta _n)$.
Hence the on-site electron energy, within the displacements' linear
approximation,  can be written as
\begin{eqnarray}
{\cal{E}}_{n} &=&{\cal{E}}_0  + \chi_1\,(U_n+U_{n-1}) + \chi_2 A_n
\nonumber\\
 &=&{\cal{E}}_0  + (\chi_1\cos \frac{\alpha }{2}+\chi
_2\sin \frac{\alpha }{2})(u_{n+1}-u_{n-1})
+2(\chi_1 \sin \frac{\alpha}{2}+\chi_2\cos\frac{\alpha }{2})s_n\nonumber\\
&&+(\chi_1 \sin \frac{\alpha}{2}-\chi_2\cos\frac{\alpha}{2})
   (s_{n+1}+s_{n-1}),
\label{en2}
\end{eqnarray}
where
$\chi _1=d V_{n\pm 1,n}(a)/da$, $\chi _2=(1/a)dV_n(\beta _n)/d\beta_n
|_{\beta _n=\alpha }$. The parameters $\chi _1$, $\chi _2$ and $G_2$
determine the interaction of the quasiparticle with the chain deformation
(exciton/electron-phonon coupling) in the linear approximation relative to 
the site displacements from their equilibrium positions.

Due to the symmetry we see that $V_n(\beta_n) = V_n(-\beta _n)$,
and we can set $V_n(\beta_n) = V_n(\cos(\beta _n))$. Therefore,
\begin{equation}
\chi_2=(1/a){dV_n(\beta _n) \over d\beta _n}|_{\beta _n=\alpha } =
 -(1/a)V_n'\sin(\alpha).
\end{equation}
Thus, $\chi_2 \sim \sin(\alpha)$ and $\chi _2 = 0$ at $\alpha = 0$ and in the
linear chain the expansion of $V_n(\beta_n)$ in series of $\Delta \beta_n$
begins with the quadratic term while the linear term vanishes.

The Hamiltonian of the site displacements, $H_{ph}$, consists of the
kinetic and potential energies of displacements. The potential energy 
consists of the potential energies of the pairwise
interactions, which depend on the distance between the sites, and on
the bending contribution. Therefore, the total Hamiltonian of the
phonons takes the form:
\begin{equation}
H_{ph} = \frac{1}{2} \sum_{n} \Bigl(
{\vec{p}_{n}^2 \over M}\,\,+\,\kappa\,(| \vec{R}_{n+1} - \vec{R}_n | - a_0)^2
+\,\kappa_b a_0^2(\beta_n - \beta_0 )^2 \Bigr) .
\label{ham-phon}
\end{equation}
Here $M$ is the mass of a lattice site, $\vec{p}_{n}$ are the momentum 
operators
canonically conjugate to the operators of displacements $\vec{r}_n$,
$ \kappa $  is the elasticity coefficient due to
the change of distance between the molecules from their equilibrium value 
$a_0$, and $ \kappa_b $ is the elasticity constant responsible for the bending 
of the bond (for small bending angles $\beta_n$).

 As our chain is closed to form a circle, we have
$\beta_n = \alpha + \Delta \beta _n = \alpha + a A_n $. From the definition
of $A_n$ we see that $\sum_n A_n = 0$ and, therefore, the equilibrium
distance between the lattice sites on the circle remains the same:  
$a = a_0$. Therefore,
the total Hamiltonian for phonons in a circle takes the form:
\begin{eqnarray}
H_{ph} &=& \frac{1}{2} \sum_{n} \Bigl(
{p_{n}^2\over M}\,+\,{q_{n}^2\over M}\,
+\,\kappa\,[\cos(\frac{\alpha}{2}) (\ur-\uc)+\sin(\frac{\alpha}{2})
(\ssc+\ssr)]^2
\nonumber\\
&&+\,\kappa_b(\sin\frac{\alpha}{2} (u_{n+1}-u_{n-1}) + \cos\frac{\alpha}{2}
(2s_n - s_{n+1} - s_{n-1}))^2\,\Bigr) + {\cal W }_b \nonumber\\
&=&  \frac{1}{2} \sum_{n} \Bigl(
{p_{n}^2\over M}\,+\,{q_{n}^2\over M}\,
+\,\kappa\,U_n^2 + \kappa_b \, A_n^2 \Bigl) + {\cal W }_b, 
\label{phon-ham}
\end{eqnarray}
where $p_{n}$ and $q_{n}$ are the momentum operators canonically conjugate
to the operators of the tangential and radial components of the site
displacements, respectively, and ${\cal W }_b = (1/2)\kappa_b a^2 \alpha^2 N =
2 \pi^2 \kappa_b a^2/N $ is the energy required to bend the linear chain into
a circle (which we will drop from now on as it is a constant).

Note that both $U_n$ and $A_n$ are invariant under the global translations of 
the circle and so is the Hamiltonian (\ref{phon-ham}), as one would expect.

Thus, the total Hamiltonian is given by
\begin{equation}
H = H_{e} + H_{ph} + H_{int},
\end{equation}
where
\begin{equation}
H_{e} = \sum_{n} \Bigl({\cal {E}}_0\, B^{\dag}_n B_n
-\,J\,(B^{\dag}_n B_{n+1} + B^{\dag}_{n+1} B_n )\Bigr) ,
\label{ham-exc}
\end{equation}
\begin{equation}
H_{int} = \sum_{n} \Bigl(
 \, B^{\dag}_n B_n [\chi_1\, (U_n+U_{n-1})+\chi_2\,A_n]
 + G_2 \,(B^{\dag}_n B_{n+1} + B^{\dag}_{n+1} B_n )\,U_n \Bigr) ,
\label{ham-exc2}
\end{equation}
and $H_{ph}$ is given by \Ref{phon-ham}.

These expressions are written in the site representation.
We can transform them into the quasimomentum representation by performing
 the unitary transformation
\begin{equation}
B_n\ =\ \frac{1}{\sqrt{N}}\sum_{k}e^{ikn}B_k ,
\label{transfB}
\end{equation}
\begin{eqnarray}
u_n\ =\ \frac{1}{\sqrt{MN}}\sum_{q}e^{iqn}U_{l,q}, \quad
p_{n}\ =\ \sqrt{\frac{M}{N}}\sum_{q}e^{-iqn}\Pi_{l,q} , \nonumber\\
s_n\ =\ \frac{1}{\sqrt{MN}}\sum_{q}e^{iqn}U_{r,q}, \quad
q_{n}\ =\ \sqrt{\frac{M}{N}}\sum_{q}e^{-iqn}\Pi_{r,q} ,
\label{transf1}
\end{eqnarray}
where $k \textrm{(\ or $q$)} = 2\pi \nu/N$, $\nu = 0, \pm1, ...,\pm(N/2-1),N/2$
for $N$ even and $\nu = 0,\pm1, ...,\pm(N-1)/2$ for $N$ odd.
This allows us to rewrite the quasiparticle Hamiltonian \Ref{ham-exc} 
and the phonon Hamiltonian \Ref{phon-ham} as
\begin{equation}
H_{e}=  \sum_{k} {\cal E}(k) B^{\dag}_k B_k, 
\label{exc-ham2}
\end{equation}
\begin{equation}
H_{ph}= \frac{1}{2} \sum_{q} \Bigl( \sum_{\mu}
(\Pi^{\dag}_{\mu,q}\Pi_{\mu,q} + \omega^2_{\mu}(q) U^{\dag}_{\mu,q}U_{\mu,q})
+ if(q) (U^{\dag}_{r,q}U_{l,q} - U^{\dag}_{l,q}U_{r,q}),
\Bigr)
\label{phon-ham2}
\end{equation}
where $\mu = l,r $, and 
\begin{equation}
{\cal E}(k) = {\cal {E}}_0\,- \,2J\,\cos(k) ,
\label{Eexc}
\end{equation}
\begin{eqnarray}
\omega^2_l(q)\,=\,4\frac{\kappa}{M}
\sin^2(\frac{q}{2})[\cos^2(\frac{\alpha}{2}) +
4c\sin^2(\frac{\alpha}{2})\cos^2(\frac{q}{2})],\,\nonumber\\
\omega^2_r(q)\,=\,4\frac{\kappa}{M}
[\sin^2(\frac{\alpha}{2})\cos^2(\frac{q}{2}) +
4c\cos^2(\frac{\alpha}{2})\sin^4(\frac{q}{2})], \,\nonumber\\
f(q)\,=\,\frac{\kappa}{M} \sin(\alpha) \sin(q) [1 +
4c\sin^2(\frac{q}{2})].
\label{omegas-f}
\end{eqnarray}
Here $c = \kappa_b/\kappa $.

Applying the unitary canonical transformation
\begin{eqnarray}
U_{l,q} = \cos(\theta (q))Q_{1,q} + i\sin(\theta (q))Q_{2,q} ,\quad
\Pi_{l,q} = \cos(\theta (q))P_{1,q} - i\sin(\theta (q))P_{2,q} , \nonumber\\
U_{r,q} = i\sin(\theta (q))Q_{1,q} + \cos(\theta (q))Q_{2,q}, \quad
\Pi_{r,q} = -i\sin(\theta (q))P_{1,q} + \cos(\theta (q))P_{2,q} ,\nonumber\\
\label{transf2}
\end{eqnarray}
in which
\begin{equation}
\tan(2\theta (q))\,=\, \frac{2f(q)}{\Delta (q)}\,, \qquad
\Delta (q)= \omega^2_l(q) - \omega^2_r(q), 
\label{thetaq}
\end{equation}
we obtain the phonon Hamiltonian in the diagonal form
\begin{equation}
H_{ph}= \frac{1}{2} \sum_{j,q} \Bigl(
P^{\dag}_{j,q}P_{j,q} + \Omega^2_j(q) Q^{\dag}_{j,q}Q_{j,q}] \Bigr),
\label{phon-ham3}
\end{equation}
where $j = 1,2$ and
\begin{equation}
\Omega^2_{j}(q) = \frac{1}{2}\Bigl(\omega^2_{l}(q) +\omega^2_{r}(q) -
(-1)^j [(\omega^2_l(q) - \omega^2_r(q)) \cos(2\theta (q)) +
2 f(q) \sin(2\theta (q))] \Bigr) .
\label{Omegaj2}
\end{equation}

Then we can introduce the operators of creation $b^{\dag}_{q,j}$ and
annihilation $b_{q,j}$ of phonons by performing a further canonical 
transformation
\begin{equation} 
Q_{j,q}= \sqrt{\frac{\hbar}{2 \Omega_{j}(q)}} (b_{q,j} + b^{\dag}_{-q,j}),
\end{equation}
\begin{equation}
 P_{j,q}= -i\sqrt{\frac{\hbar \Omega_{j}(q)}{2}} (b_{q,j} - b^{\dag}_{-q,j}),
\label{transf3}
\end{equation}
and find that the total Hamiltonian can be rewritten as
\begin{equation}
H = \sum_{k}{\cal E}(k) B^{\dag}_k B_k +
 \sum_{j,q} \hbar \Omega_j(q)\Bigl( b^{\dag}_{q,j}b_{q,j} +
\frac{1}{2} \Bigr) + 
\end{equation}
\begin{equation}
+ \frac{1}{\sqrt{MN}}
 \sum_{k,q,j}\Phi_j(k,q) B^{\dag}_{k+q} B_k (b_{q,j} + b^{\dag}_{-q,j} )
\label{Htot}
\end{equation}
where the coupling functions $\Phi_j(k,q)$ are given by
\begin{equation} 
\Phi_j(k,q) = F_j(k,q) \sqrt{\frac{\hbar}{2 \Omega_{j}(q)}} 
\end{equation}
with
\begin{eqnarray}
F_{1}(k,q) &=& 4 i [F_l(k,q) \cos(\theta (q)) + F_r(k,q)\sin(\theta (q))],
\nonumber\\
F_{2}(k,q) &=& 4 [F_r(k,q) \cos(\theta (q)) - F_l(k,q)\sin(\theta (q))].
\label{Gs}
\end{eqnarray}
Here
\begin{eqnarray}
F_l(k,q)\ = \ [(\chi_1 \cos(\frac{\alpha}{2}) +
\chi_2 \sin(\frac{\alpha}{2})) \cos(\frac{q}{2}) +
G_2 \cos(\frac{\alpha}{2}) \cos(k+\frac{q}{2})]
\sin(\frac{q}{2}) ,\,\nonumber\\
F_r(k,q) = \chi_1 \sin(\frac{\alpha}{2})\cos^2(\frac{q}{2}) +
\chi_2 \cos(\frac{\alpha}{2})\sin^2(\frac{q}{2}) +
G_2 \sin(\frac{\alpha}{2}) \cos(k+\frac{q}{2}) \cos(\frac{q}{2}) .\nonumber\\
\label{phy-varphi}
\end{eqnarray}
Note that $\Phi^*_j(k,q) = \Phi_j(k+q,-q)$.

At large values of $N$ the parameter  $\alpha$ is small ($N \to \infty$, 
$\alpha \to 0$).
At $\alpha = 0$  we have a linear molecular chain. In this case, according to
\Ref{omegas-f}, $f(q) = 0$ and the longitudinal $Q_{1,q} = U_{l,q}$ (along
the chain axis) and the transversal $Q_{2,q} = U_{r,q}$ displacements do not
mix ($\theta (q) = 0$ in \Ref{transf2}). A linear chain is characterised
by the longitudinal acoustic waves $U_{l,q}$ with frequency
\begin{equation}
\omega^2_l(q)\,=\,4\frac{\kappa}{M} \sin^2(\frac{q}{2}),
\label{acoustphon}
\end{equation}
and the bending waves $U_{r,q}$ with frequency
\begin{equation}
\omega^2_r(q)\, = \,16 \frac{\kappa_b}{M} \sin^4(\frac{q}{2}).
\label{bendingphon}
\end{equation}

In a linear chain in the linear approximation with regard 
to the displacements, quasiparticles interact 
 only with the longitudinal acoustic phonons because 
\begin{equation}
F_1(k,q) = 4iF_l(k,q) = 4i [(\chi_1 \cos(\frac{q}{2}) +
G_2 \cos(k+\frac{q}{2})] \sin(\frac{q}{2})
\end{equation}
and $F_2(k,q) = 4F_r(kq) = 4 \chi_2 \sin^2(\frac{q}{2}) = 0$ (as
$\chi_2 \sim \sin(\alpha)$ and so $\chi _2 = 0$ at $\alpha = 0$).

Note that in a long linear chain, $N$ which is included in the definition of
the one-dimensional quasimomentum $q$, determines the main region of
quantization. So, we have
$q = 2\pi \nu/N =\alpha \nu$, $\nu = 0, \pm1, ...$.
On a circle the longitudinal (tangential) and the transversal (radial) 
displacements are mixed in a way determined by the angle 
$\theta (q)$. From the definition
\Ref{thetaq} we see that $\tan(2\theta (q)) = - \tan(2\theta (-q))$. This
means that $2\theta (q) = - 2\theta (-q)$ or $2\theta (q) = \pi -
2\theta (-q)$. We choose the first relation and, consequently, 
$\theta (0) = 0$ and $-\pi/2 \leq 2\theta (q) \leq \pi/2$.

According to \Ref{Omegaj2}, in a circle with $N$
sites, the phonon Hamiltonian has two normal branches with different
$q$ which take $N$ values, i.e. $2N$ frequencies. Note that 
in a circle with $N$ particles, each of which has two 
degrees of freedom, the number of normal vibrations is
$2N-3$ because there are two degrees of freedom which correspond to the 
motion of the
centre of mass of the system and one degree corresponds to the rotation
of the circle. And indeed, there are three zero frequencies 
in the vibrational spectrum \Ref{Omegaj2}, namely: $\Omega_{1}(0) =
\Omega_{2}(\pm 2\pi/N) = 0 $. At $q=0$, $\theta (0) = 0$ and the normal
coordinate $Q_{1,q=0}$ corresponds
to the site displacements $u_n = 1/\sqrt{MN} Q_{1,0}$ and $s_n = 0$, i.e. this
is the rotation of the circle. At $q=\pm 2\pi/N = \pm\alpha$
$\theta (\pm\alpha) = \pm\pi/4$, the normal
coordinates $Q_{2,\pm\alpha}$ describe the site displacements: $\vec{r}_n =
\vec{e}_x \sqrt{2/(MN)} Q_{im2,\alpha}$ and $\vec{r}_n =
\vec{e}_y \sqrt{2/(MN)} Q_{re2,\alpha}$ and correspond to the motion of
the centre of mass in two mutually orthogonal directions. Here we introduce
the notations
$Q_{rej,\alpha} = [Q_{j,\alpha} + Q^*_{j,\alpha}]/2$ and
$Q_{imj,\alpha} = i[ Q_{j,\alpha} - Q^*_{j,\alpha}]/2$ after taking into 
account the relation $ Q_{j,-\alpha} = Q^*_{j,\alpha}$. It is natural that
the quasiparticles do not interact with these normal coordinates. From \Ref{Gs}
and \Ref{phy-varphi} we see that $F_1(k,0) = F_2(k,\pm\alpha) = 0$.

For the values $q=0$ and $q=\pm\alpha$ we have two nonzero frequencies which
have no analogue in an open chain. The frequency
\begin{equation}
\Omega^2_{2}(0)\,=\,4\frac{\kappa}{M} \sin^2(\frac{\alpha}{2})
\label{Omegabr}
\end{equation}
is the frequency of the total symmetrical vibration which corresponds to the
circle breathing, i.e. to the deviation of the circle radius from its 
equilibrium
value $R_0$, because the normal coordinate $Q_{2,q=0}$ describes the
site displacements $s_n = 1/\sqrt{MN} Q_{2,0}$ and $u_n = 0$. The 
doubly degenerate frequency
\begin{equation}
\Omega^2_{1}(\pm\alpha)\,=\,8\frac{\kappa}{M}
\sin^2(\frac{\alpha}{2})\cos^2(\frac{\alpha}{2}) [1 +
4c\sin^2(\frac{\alpha}{2})]
\label{Omegadef}
\end{equation}
is the frequency of vibrations which correspond to the nonsymmetrical 
deformation of the circle  into two mutually orthogonal directions.
The normal coordinates $Q_{1,\pm\alpha}$ describe the site displacements
$\vec{r}^{(j)}_n = \vec{e}^{(j)}_n A_{j}$ ($j=1,2$) which are
characterised by the two mutually orthogonal unit vectors of deformation
\begin{eqnarray}
\vec{e}_{n}^{(1)} &=& \vec{e}_x \cos(2n\alpha + \theta_0) - \vec{e}_y
\sin(2n \alpha + \theta_0), \\
\vec{e}_{n}^{(2)} &=& \vec{e}_x \sin(2n\alpha + \theta_0) + \vec{e}_y
\cos(2n \alpha + \theta_0),
\end{eqnarray}
and the amplitudes $A_{1} = \sqrt{2/(MN)} Q_{re1,\alpha}$, 
$A_{2} = \sqrt{2/(MN)} Q_{im1,\alpha}$. The interaction of the quasiparticles
with these normal coordinates is determined by the coupling functions
\begin{eqnarray}
F_{2}(k,0) &=& 4 \sin(\frac{\alpha}{2}) (\chi_1 + G_2 \cos(k) ) ,\\
F_1(k,\pm\alpha)\ &=& \ \pm 2\sqrt{2}i \sin(\alpha)
[\chi_1 \cos(\frac{\alpha}{2}) + \chi_2 \sin(\frac{\alpha}{2}) +
G_2 \cos(k\pm\frac{\alpha}{2})] .\nonumber\\
\label{Gs0alph}
\end{eqnarray}

For other frequencies with $q > \alpha$ we have
\begin{equation}
\Omega^2_{1,2}(q) = \frac{1}{2}(\omega^2_{1} +\omega^2_{2} \mp
\sqrt{\Delta ^2 + 4f^2}), 
\label{Omega2}
\end{equation}
and the electron-phonon coupling functions $F_i$ (see \Ref{Gs}) in which 
\begin{equation}
\cos(\theta (q)) = \frac{1}{\sqrt{2}}\sqrt{1+\frac{\Delta}{
\sqrt{\Delta ^2 + 4f^2}} },
\end{equation}
\begin{equation}
\sin(\theta (q)) = \frac{\sqrt{2}f}
{\sqrt{(\sqrt{\Delta ^2 + 4f^2}+\Delta)\sqrt{\Delta ^2 + 4f^2}}},
\label{a-b}
\end{equation}

\section{Equations in the adiabatic approximation}

Next we consider the ground state of the circle with
one quasiparticle. We assume that the
adiabatic approximation is applicable and that we can choose the
wave-function of the ground state in the form
\begin{equation}
|\psi\rangle = U|\psi_e\rangle ,
\label{adappr}
\end{equation}
where $U$ is the unitary operator of the coherent molecule
displacements, 
\begin{equation}
|\psi_e(t)\rangle=
\sum_{n}\phi_{n}(t)B\sp{\dagger}_{n}|0\rangle =
\sum_{k}\psi(k,t)B\sp{\dagger}_{k}|0\rangle ,
\label{grf}
\end{equation}
where the functions $\phi _{n}$ and $\psi(k)$  satisfy the normalisation
condition:
\begin{equation}
\sum_{n} |\phi _{n}|^2=1,  \quad
\sum_{k} |\psi (k)|^2=1.
\label{norm}
\end{equation}
Here the functions $\phi _{n}$ and $\psi(k)$ are connected by the same unitary
transformation from the site representation to the quasimomentum one,
\Ref{transfB}, as the operators $B_n$ and $B_k$.

The zero-order adiabatic approximation 
is equivalent to the semi-classical approach when the
molecule displacements and their canonically conjugate momenta are
considered as classical ones. Indeed, by computing the mean value of the
Hamiltonian over the state \Ref{adappr}, which is a functional of
$\phi _n$, and of the molecule displacements and of their canonically 
conjugate momenta,
we transform the problem to a classical one which involves
classical variables instead of operators. In the site representation
the Hamiltonian functional then reads as
\begin{eqnarray}
{\cal H} &=& \sum_{n} \Bigl(
{\cal {E}}_0\,\cmod{\ph}\,
-\,j\,(\ph^*\phr+\phr^*\ph)
 +\chi_1\cmod{\ph}(U_n+U_{n-1})+\chi_2\,\cmod{\ph}A_n\nonumber\\
 &+& G_2(\ph^*\phr+\phr^*\ph)U_n
\Bigr)\,+\,{\cal W },
\label{Hamiltonian}
\end{eqnarray}
where all the variables are c-numbers and ${\cal W}$ is the mean value of the
phonon Hamiltonian, $H_{ph}$, {\it i.e.} of the classical Hamiltonian of 
harmonic
vibrations of the circle. We can also represent this Hamiltonian in the
quasimomentum representation
\begin{eqnarray}
{\cal H} &=& \sum_{k}{\cal E}(k) \,\cmod{\psi(k)}\,  +
 \sum_{j,q} \hbar \Omega_j(q) \cmod{\beta_{q,j}} \nonumber\\
 &+& \frac{1}{\sqrt{MN}}
 \sum_{k,q,j}\Phi_j(k,q) \psi^{*}(k+q) \psi(k) (\beta_{q,j} +
\beta^{*}_{-q,j} ),
\label{Ham-qm}
\end{eqnarray}
where the variables $\psi(k)$ and $\beta_{q,j}$ are connected to
$\phi _n$ and the molecule displacements $\uc$ and $\ssc$ and their
canonically conjugate momenta $p_{n}$ and $q_{n}$ by the same 
canonical unitary transformations \Ref{transfB}, \Ref{transf1}, \Ref{transf2},
\Ref{transf3} as the corresponding operators.

Then we can write down the Hamiltonian equations of motion for these variables.
In particular, we get
\begin{eqnarray}
p_{n}\,&=&\,M {d\uc\over dt}, \nonumber\\
q_{n}\,&=&\,M {d\ssc\over dt}.
\label{p-q}
\end{eqnarray}

We will restrict our attention to the stationary solutions for which 
$\phi _n (t) = \exp(-iEt/\hbar)\phi _n $ 
(or $\psi (k,t) = \exp(-iEt/\hbar)\psi (k)) $
and $p_n = q_n = 0$. Performing the variations of (\ref{Hamiltonian})
with respect to the variables $\ph$, $u_i$ and $s_i$, we obtain the 
static equations 
\begin{eqnarray}
0 &=& ({\cal {E}}_0\,-\,E)\ph - \,J\,(\phr+\phl)
+\chi_1 \ph\,(U_i+U_{i-1})+\chi_2 \ph\,A_i \nonumber\\
&+&G_2(\phr\, U_i\,+\,\phl\,U_{i-1}), \\
\label{Eqnphi}
0 &=& \kappa (\cos^2(\frac{\alpha}{2})(2\,\uc-\ur-\ul)+
         \cos(\frac{\alpha}{2})\sin(\frac{\alpha}{2})(\ssl-\ssr) )\nonumber\\
     &+& \kappa_b (\sin^2(\frac{\alpha}{2})(2\,\uc-\urr-\ull)+
       \cos(\frac{\alpha}{2})\sin(\frac{\alpha}{2})(2\ssl-2\ssr +\ssrr-\ssll)) 
\nonumber\\
       &+&(\chi_1\cos(\frac{\alpha}{2})+\chi_2\sin(\frac{\alpha}{2}))
          \,(\cmod{\phl}-\cmod{\phr})\nonumber\\
       &+&G_2\,\cos(\frac{\alpha}{2})(\phl^*\ph+\ph^*\phl-\phr^*\ph-\ph^*\phr),
\label{Eqnu}
\\
0 &=& \kappa \sin(\frac{\alpha}{2}) (\sin(\frac{\alpha}{2})(2\ssc +\ssr+\ssl)
         +\cos(\frac{\alpha}{2}) \,(\ur-\ul))
\nonumber\\
&+&\kappa_b\,[\cos^2(\frac{\alpha}{2})
   (6\ssc -4 \ssr -4 \ssl +\ssrr + \ssll) \nonumber\\
&& + \cos(\frac{\alpha}{2})\sin(\frac{\alpha}{2})(2\ur - 2\ul + \ull -\urr)]
\nonumber\\
&+&2(\chi_1\sin(\frac{\alpha}{2})+\chi_2\cos(\frac{\alpha}{2}))\cmod{\ph}
+(\chi_1\sin(\frac{\alpha}{2})
-\chi_2\cos(\frac{\alpha}{2}))(\cmod{\phr}+\cmod{\phl})
\nonumber\\
  &+&G_2\sin(\frac{\alpha}{2})(\phl^*\ph+\ph^*\phl+\phr^*\ph+\ph^*\phr).
  \label{Eqna}
\end{eqnarray}
Note that the equations (\ref{Eqnphi}-\ref{Eqna}) are invariant
under the following transformation:
\begin{eqnarray}
u_i &\rightarrow& -u_i, \nonumber\\
s_i &\rightarrow& -s_i, \nonumber\\
\chi_n &\rightarrow& -\chi_n  \,\,\, (n = 1,2),  \nonumber\\
G_2 &\rightarrow& -G_2. 
\label{Symmetry}
\end{eqnarray}

In the quasimomentum representation this system of equations can 
be rewritten in the following form
\begin{equation}
({\cal E}(k) - E)\,\psi(k)\, + \frac{1}{\sqrt{MN}}
 \sum_{q,j}F_j(k-q,q) Q_j (q) \psi(k-q)\, = \,0 ,
\label{Eqpsi}
\end{equation}
\begin{equation}
\Omega^2_j(q) Q_j (q) = - \frac{1}{\sqrt{MN}}
 \sum_{k}F^*_j(k,q) \psi^{*}(k) \psi(k+q),
\label{EqQj}
\end{equation}
where
\begin{equation}
Q_{j}(q) = \sqrt{\frac{\hbar}{2 \Omega_{j}(q)}} (\beta_{q,j} +
\beta^{*}_{-q,j}) .
\end{equation}

From Eq. (\ref{EqQj}) we obtain the explicit expressions for
$Q_{j}(q)$, namely:
\begin{equation}
Q_j (q) = - \frac{1}{\sqrt{MN}}
 \sum_{k} \frac{F^*_j(k,q)}{\Omega^2_j(q)} \psi^{*}(k) \psi(k+q).
\label{Qjq}
\end{equation}

Substituting these expressions into (\ref{Eqpsi}), we obtain the
following nonlinear integral equation for $\psi(k)$
\begin{eqnarray}
&&({\cal E}(k) - E)\,\psi(k)\, -
\frac{1}{N} \sum_{q,k_1(|q| > \alpha)}G(k,k_1,q)
\psi^{*}(k_1) \psi(k_1+q) \psi(k-q)\, \nonumber\\
&-&\frac{1}{N} \sum_{k_1} {\cal F}_0(k,k_1)
\psi^{*}(k_1) \psi(k_1) \psi(k)\,\nonumber\\
&-&\frac{1}{N} \sum_{k_1,q=\pm \alpha} {\cal F}_{\alpha}(k,k_1,q)
 \psi^{*}(k_1) \psi(k_1+q) \psi(k-q)  = \,0 ,
\label{NLEqpsik}
\end{eqnarray}
where
\begin{eqnarray}
\label{Gkkq}
G (k,k_1,q)\,&=&\,\sum_{j} \frac{F_j(k-q,q)F^*_j(k_1,q)}{M\Omega^2_j(q)}, \\
\label{F0}
{\cal F}_0(k,k_1)\, &=&\, \frac{F_2(k,0)F^*_2(k_1,0)}{M\Omega^2_2(0)}, \\
\label{Falpha}
{\cal F}_{\alpha}(k,k_1,\alpha)\,&=&
\frac{F_1(k-\alpha,\alpha)F^*_1(k_1,\alpha)}{M\Omega^2_1(\alpha)} .
\end{eqnarray}
Taking into account expressions (\ref{Omegaj2}) and (\ref{Gs}), and
then Eqs.(\ref{omegas-f}) and (\ref{phy-varphi}), we obtain
\begin{equation}
G(k,k_1,q) = \frac{{\chi_2}^2}{\kappa_b} + G_1(k,k_1,q),
\end{equation}
\begin{equation}
G_1(k,k_1,q) = \frac{4}{\kappa }\{ \chi_1^2 \cos^2(\frac{q}{2}) +
\chi_1 G_2 [\cos(k-\frac{q}{2})+ \cos(k_1+\frac{q}{2})]\label{Gexpl}
\end{equation}
$$ +G_2^2 \cos(k-\frac{q}{2})\cos(k_1+\frac{q}{2})\}.$$
Substituting expressions (\ref{Omegabr}), (\ref{Omegadef}) and (\ref{Gs0alph})
into (\ref{F0}) and (\ref{Falpha}), we get 
\begin{eqnarray}
{\cal F}_0(k,k_1)\, &=&\,G_1(k,k_1,0) ,
 \nonumber\\
{\cal F}_{\alpha}(k,k_1,\alpha)\,&=&\, G_1(k.k_1,\alpha)+
\frac{4}{\kappa} \sin(\frac{\alpha}{4}) f(k,k_1,\alpha),
\label{F2-F1}
\end{eqnarray}
where $f(k,k_1,)$ is a function of $k$ and $\alpha$,  the explicit expression 
of which will be not essential for us.

\section{Fully delocalised states}

Equations (\ref{Eqnphi}-\ref{Eqna}) or (\ref{Eqpsi}-\ref{EqQj}) always
admit a simple solution for all values of the parameters that 
correspond to a fully delocalised quasiparticle state, {\it i.e.} for which
the quasiparticle is evenly distributed along all the
chain: $\ph = 1/(\sqrt{N})\exp(ik_0 n)$. In the $k$-representation this
solution is fully localised $\psi(k) = \delta_{k,k_0}$.
 From \Ref{EqQj} we see that in this case $Q_j(q) \sim
\delta_{q,0}$ and, therefore, only the normal coordinate $Q_2(0)$ is nonzero.
For $j>0$ the lowest energy state corresponds to the value $k_0=0$.
Defining
\begin{equation}
\Lambda = E\,-\,{\cal {E}}_0\,+\,2 J, 
\label{Lambda}
\end{equation}
this fully delocalised solution for the ground state is given by
\begin{eqnarray}
\psi(k) &=& \delta_{k,0} \nonumber\\
Q_{2}(0) &=& -\frac{1}{\sqrt{MN}} \frac{F^*_2(0,0)}{\Omega^2_2(0)} =
- \sqrt{\frac{M}{N}} \frac{\chi_1 + G_2}{\kappa \sin(\frac{\alpha}{2})}
\label{delocks}
\end{eqnarray}
in the $k$-representation, and
\begin{eqnarray}
\ph &=& \frac{1}{N^{1/2}}, \nonumber\\
\uc &=& 0, \nonumber\\
\ssc &=& s = -\frac{1}{N \kappa} {\chi_1+G_2 \over \sin\frac{\alpha}{2}}
  \label{delocsr}
\end{eqnarray}
in the site representation with
\begin{equation}
\Lambda = -\frac{4(\chi_1 + G_2)^2}{N \kappa} = \lambda J.
\end{equation}
According to \Ref{Lambda}, $\Lambda = \lambda J$ determines the
eigenvalue of Eq. \Ref{Eqnphi} (or \Ref{Eqpsi})
\begin{equation}
E = {\cal {E}}_0\,-\,2 J\,+\,\Lambda = {\cal E}(0)\,+\,\Lambda , 
\end{equation}
where ${\cal E}(0) = {\cal {E}}_0\,-\,2 J$ is the lowest energy level
\Ref{Eexc} of a quasiparticle in the circle.

The total energy of the system is $E_{tot} = E + {\cal W}$.
For the fully delocalised solution we have ${\cal W} =
(1/2)\sum_{j,q}\Omega^2_j(q)|Q_j (q)|^2 = (1/2)\Omega^2_2(0)|Q_2 (0)|^2 $.
So, the total energy of the delocalised state is
\begin{equation}
E_{tot} = {\cal {E}}_0\,-\,2 J\,-\,\frac{2 (\chi_1 + G_2)^2}{N \kappa} =
{\cal E}(0) - \epsilon J, 
\label{Endeloc}
\end{equation}
where $\epsilon$ characterises the decrease of energy due to the 
electron-phonon interaction. Thus, we see that
the localised state of a quasiparticle in
a circle is characterised by the following values of $\lambda$ and $\epsilon$
\begin{equation}
\lambda _{del}=-\frac{2g_l}{N}, \quad \epsilon_{del}=\frac{g_l}{N},
\label{eidel}
\end{equation}
with
\begin{equation}
g_l = \frac{2 (\chi_1 + G_2)^2}{J \kappa} .
\label{gl}
\end{equation}

In this state all the $s_n$ are the same and all the $u_n$ vanish;
{\it i.e.} the circle radius changes due to the alteration of the
distance between sites $U_n =  -\frac{2}{N \kappa} (\chi_1+G_2)$. Note that,
in the limit of large $N$, $\sin \alpha\approx 2\pi/N$ and
$\sin\frac{\alpha}{2}\approx \pi/N$.
Thus, $s_i = -(\chi_1+G_2)/(\pi \kappa)$ which is independent of
$\alpha$.

\section{Numerical Solutions}

The discrete system of equations (\ref{Eqnphi}-\ref{Eqna}) can be solved
numerically. As mentioned above, the system always possesses  a fully 
delocalised solution in which the quasiparticle is evenly distributed between 
the chain sites. 
But for most values of the parameters, this is not the solution with the 
lowest energy. In this section, we  describe the numerical solutions   
(we have fixed the following values $J=1$, $k=1$ and $k_b = 0.1$).

We have found several regimes of quasiparticle localisation.
The
quasiparticle can be strongly localised on one or more sites (Fig. 1a), 
delocalised over most of the chain like (Fig 1c). Moreover, the maximum
of $|\ph|^2$ can be located on one site (Fig 1a) or two sites  (Fig 1b). Like
all the solutions that we have found, these solutions have a reflection
symmetry, but in the first case the axis of symmetry passes through a 
circle site while in the other one, it passes in between two sites.

We would like to point out here that the actual position of the soliton on the
chain is completely arbitrary and on each of the solution presented here it
was picked up more or less randomly by the relaxation method.
Moreover, one should note that given our choice of signs (for its definition),
$\lambda$ corresponds to the binding energy and so the ground state 
is the one for which $\lambda$ is the largest.

\begin{figure}[htbp]
\unitlength1cm \hfil
\begin{picture}(12,10)
 \epsfxsize=6cm \put(0,5){\epsffile{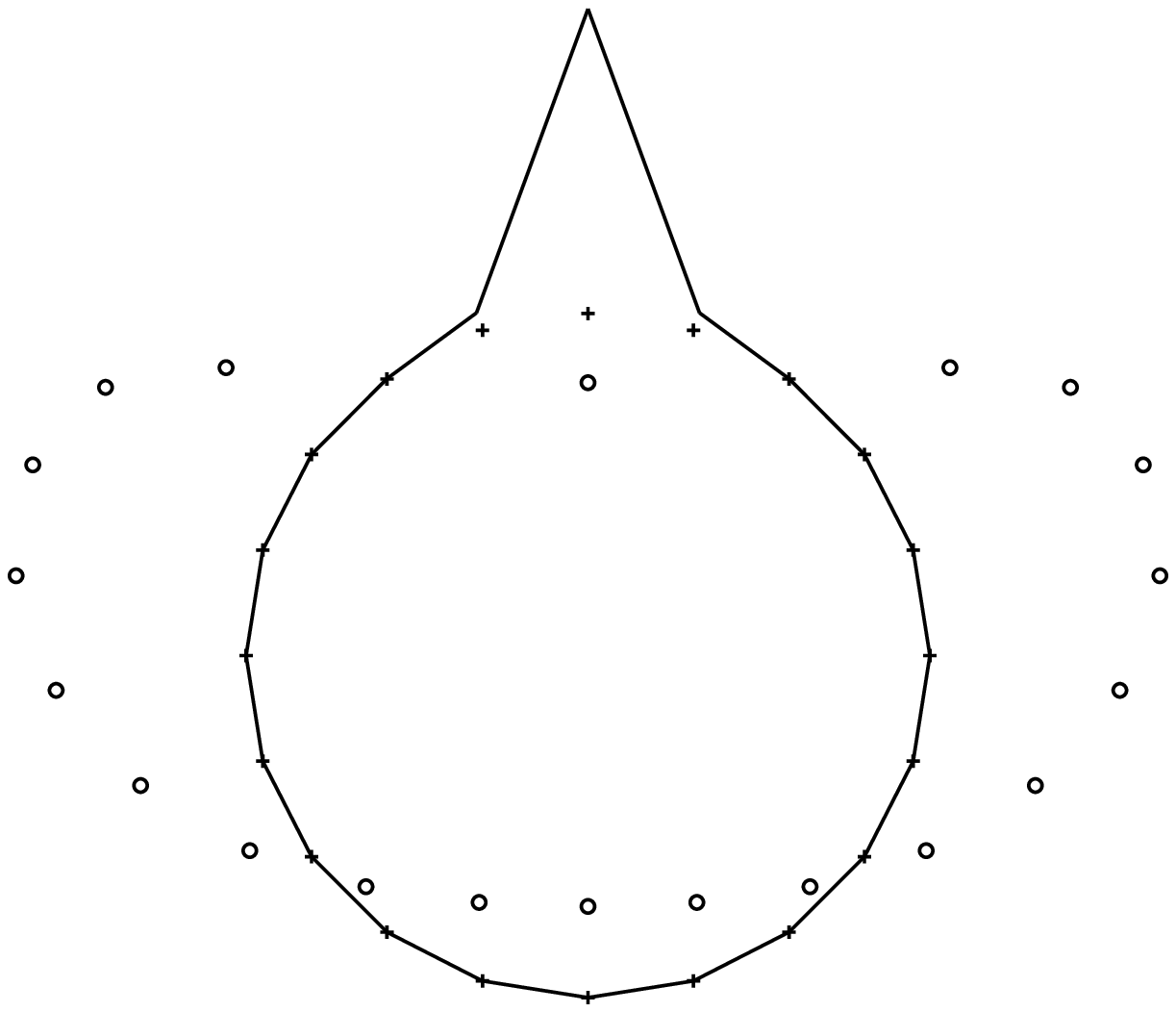}}
 \epsfxsize=6cm \put(6,5){\epsffile{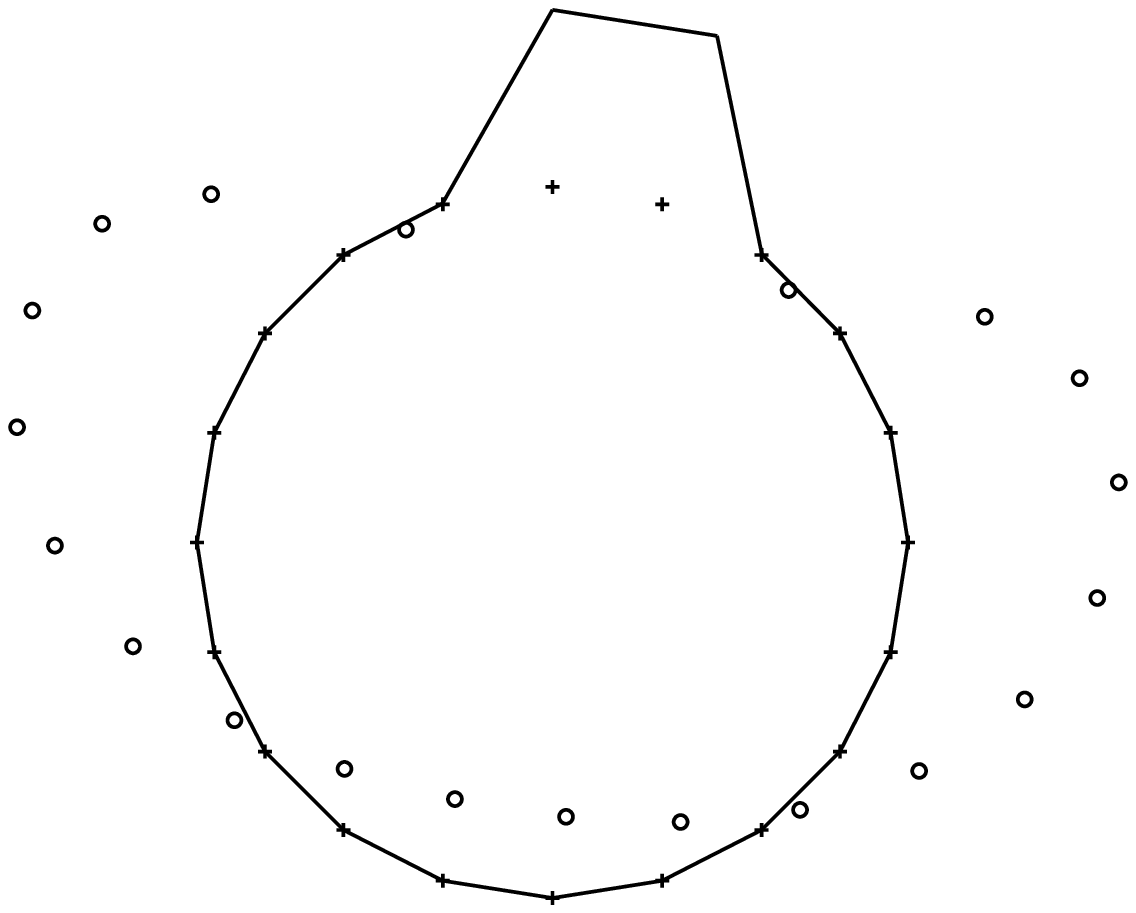}}
 \epsfxsize=6cm \put(0,0){\epsffile{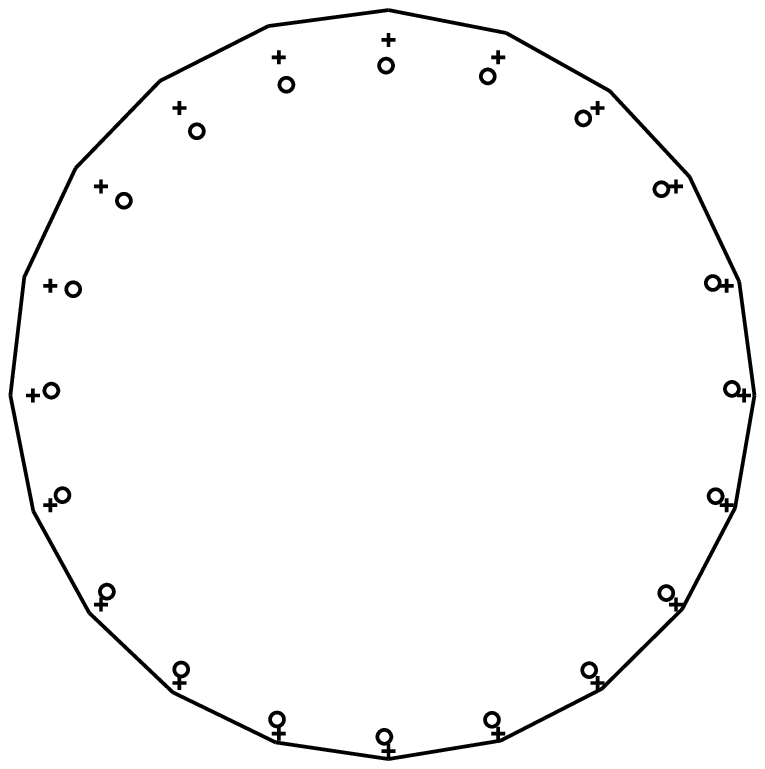}}
 \epsfxsize=6cm \put(6,0){\epsffile{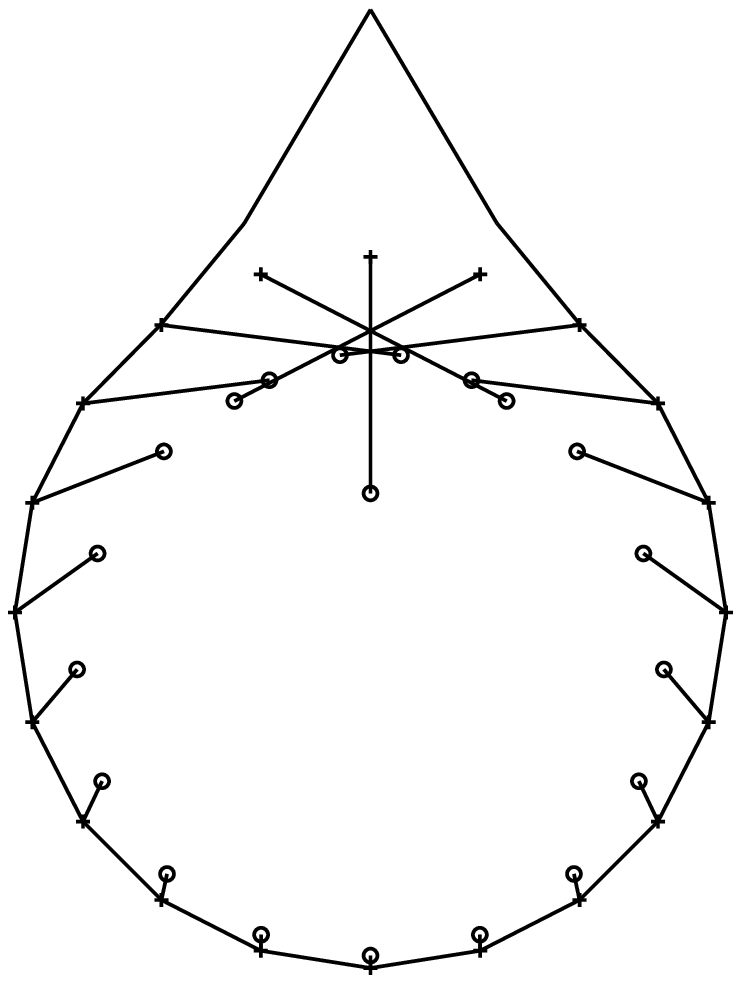}}
\put(1.5,5){a}
\put(7.5,5){b}
\put(1.5,0){c}
\put(7.5,0){d}
\end{picture}
\caption{\label{FIGsolA}
Numerical solutions of Eqs. (\ref{Eqnphi}-\ref{Eqna}) for $G_2=1$,
$\chi_2=0.2$ and a) $\chi_1=-2$ b) $\chi_1=-1.5$ c) $\chi_1=-0.5$
d) $\chi_1=2$. Symbols $+$ represent the undeformed chain. Symbols 
in the form of small circles
represent the deformed chain, while the continuous line represents $|\ph|^2$
plotted on the undeformed chain. In d), we also show where each site has been
moved to by the chain deformation.}
\end{figure}

One also notices that the chain can be deformed in different ways. For
example, in Figs. 1a and 1b, the chain is pinched into the shape of a bean 
while
in Fig. 1d the chain is compressed with a kink appearing where the soliton is
located. The deformation of the chain in all the figures presented here has
been scaled in exactly the same way to emphasise the deformation of the
chain and show the relative magnitude of the deformation when the model
parameters are varied. The actual amplitude of deformation 
depends on the values of the parameters and is much
smaller than the distance between the chain sites.

Notice also that by virtue of the symmetry \Ref{Symmetry}, the solution in 
Fig. 1.a can be transformed into a solution for $G_2=-1$, $\chi_2=-0.2$ and
$\chi_1=2$ where, as the sign of $u$ and $s$ changes,
the chain is pinched on the sides so that the lattice sites
where the soliton is located move outside the circle. By the same symmetry,
the solution shown in Fig 1.d is transformed into a ring that is
stretched out.

For large values of $\chi_1$ and $\chi_2$, the solution is usually strongly
localised while in the region $\chi_1 = -G_2$, the soliton is fully
delocalised. This is shown in Figs. 2-4 where we represent the value of the
maximum of $|\ph|^2 $ as a function of $\chi_1$ and $\chi_2$ for different
values of $G_2$.

\begin{figure}[htbp]
\unitlength1cm \hfil
\begin{picture}(8,8)
 \epsfxsize=10cm \put(0,0){\epsffile{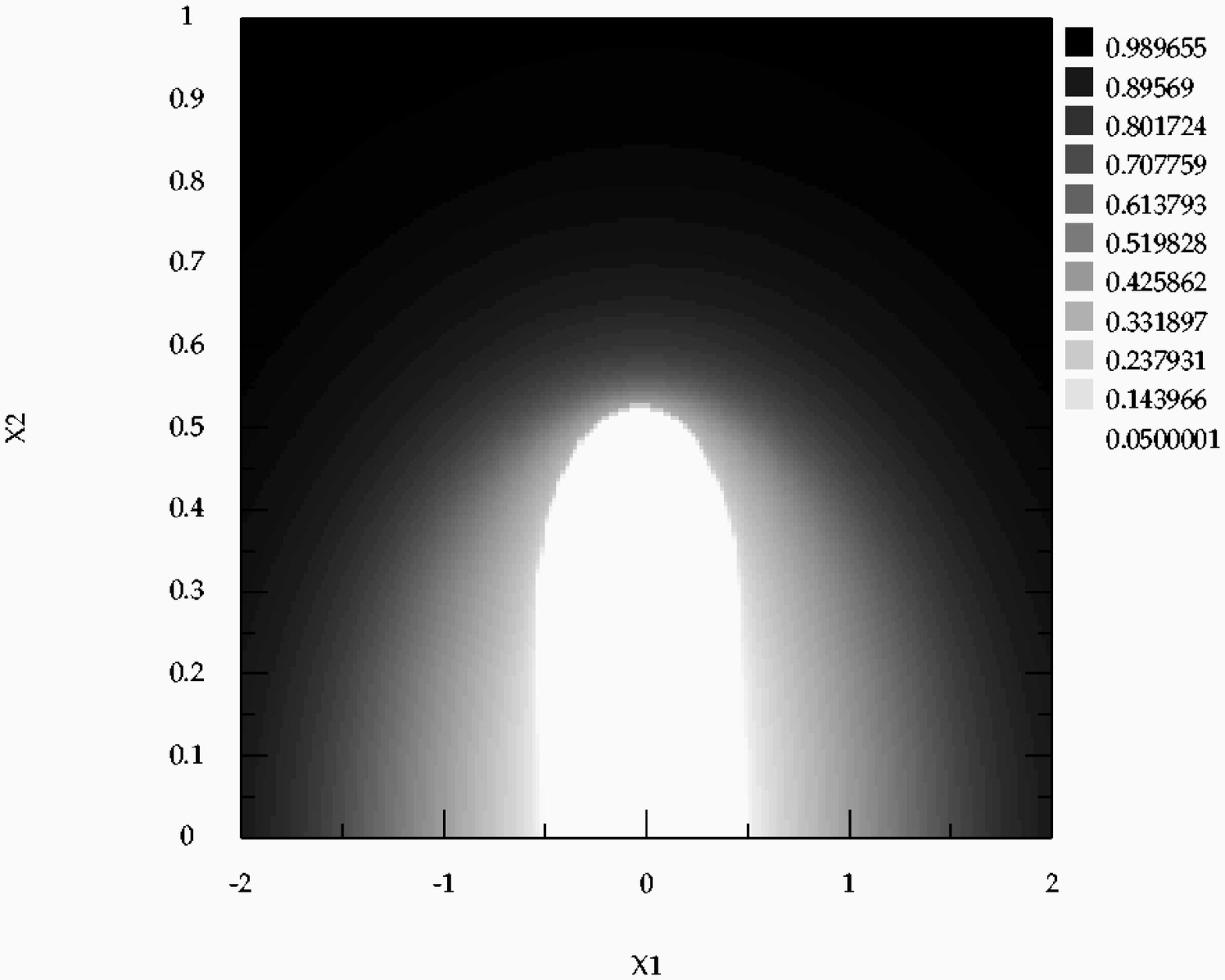}}
\end{picture}
\caption{\label{FIGplotG20}
Maximum of $|\ph|^2$ for the solutions of Eqs. (\ref{Eqnphi}-\ref{Eqna})
at $G_2=0$.} Black corresponds to $|\ph|^2 \approx 1$.
\end{figure}

When $G_2=0$, Fig \ref{FIGplotG20}, one can see that if $\chi_2 > 0.55$,
the soliton is
always strongly localised, while for smaller values, the soliton is
strongly localised  when $\chi_1 << 0$.
As $\chi_1$ increases, the quasiparticle becomes
more and more localised, then it becomes fully delocalised around
$\chi_1 = -0.5$. When $\chi_1$ is larger than $0.5$, the quasiparticle becomes
localised again and it is fully localised as $\chi_1$ goes to $\infty$.

\begin{figure}[htbp]
\unitlength1cm \hfil
\begin{picture}(8,8)
 \epsfxsize=10cm \put(0,0){\epsffile{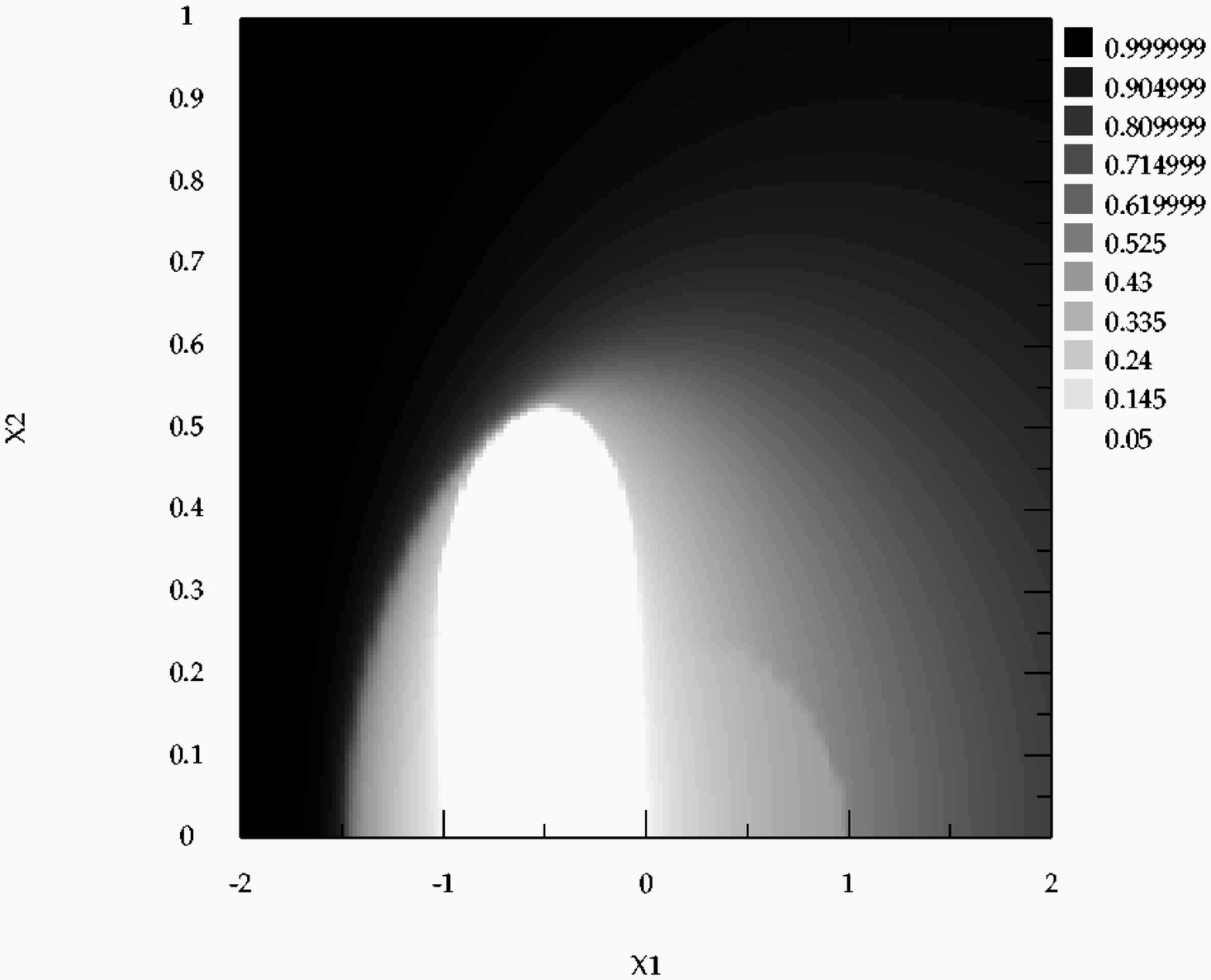}}
\end{picture}
\caption{\label{FIGplotG20.5}
Density of the solutions of Eqs. (\ref{Eqnphi}-\ref{Eqna}), $|\ph|^2$, at 
$G_2=0.5$.}
\end{figure}

At $G_2=0.5$, Fig. \ref{FIGplotG20.5}, we see a similar pattern as for $G_2=0$
except that the
region where the soliton is delocalised is centred around $\chi_1 = -G_2$.
For small value of $\chi_2$, there is also a region where the quasipartivle is
delocalised ($-1.5 < \chi_1 < -1$ when $\chi_2 \approx 0$).
The solutions in that region are delocalised but have the same symmetry as the
fully localised solitons, {\it i.e.} the axis of symmetry passes through a
chain site. In the positive range of  $\chi_1$ there is also a region where the
quasiparticle is delocalised ($0 < \chi_1 < 1$ when $\chi_2 \approx 0$) but in
this case the solution has a different symmetry, {\it i.e.}  $|\ph|^2 $  takes
its maximum value on two sites  and the axis of symmetry passes between two 
chain sites.

\begin{figure}[htbp]
\unitlength1cm \hfil
\begin{picture}(8,8)
 \epsfxsize=10cm \put(0,0){\epsffile{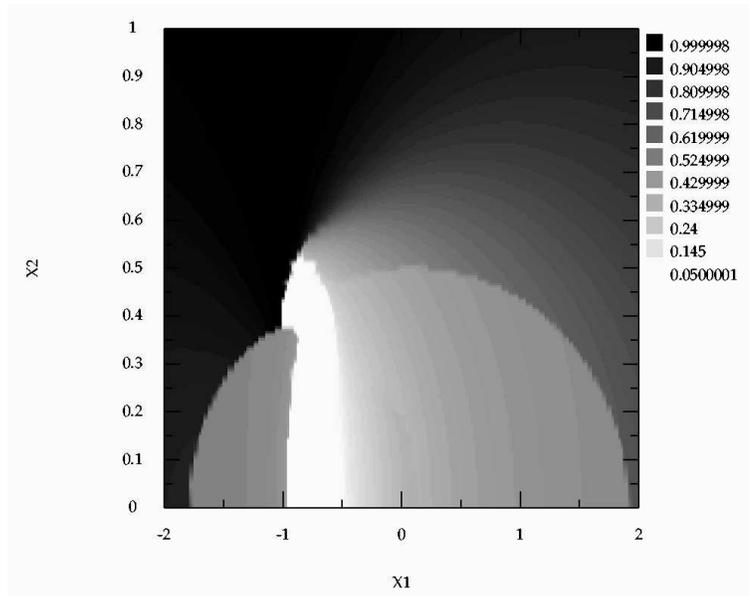}}
\end{picture}
\caption{\label{FIGplotG21}
$|\ph|^2$ density for the solutions of (\ref{Eqnphi}-\ref{Eqna})
for $G_2=1$.}
\end{figure}

When $G_2=1$, Fig \ref{FIGplotG21}, we can distinguish 4 different regions.
On the outside, the black region corresponds to fully localised solutions
(Fig. \ref{FIGsolA}a).
On the left, there is a sharp grey region ($-1.8 < \chi_1 < 1$
at $\chi_2 \approx 0$) where the
quasiparticle is mostly localised on two neighbouring sites 
(Fig. \ref{FIGsolA}b).
In the white region, the quasiparticle is fully delocalised 
(Fig. \ref{FIGsolA}c)
while in the grey half circle, on the right hand side, the quasiparticle 
is delocalised with the axis of symmetry passing between two chain sites . 
Outside
this circle the quasiparticle is localised on one site, hence, its symmetry
is different (Fig. \ref{FIGsolA}d).

\begin{figure}[htbp]
\unitlength1cm \hfil
\begin{picture}(8,12)
 \epsfxsize=8cm \put(-4,0){\epsffile{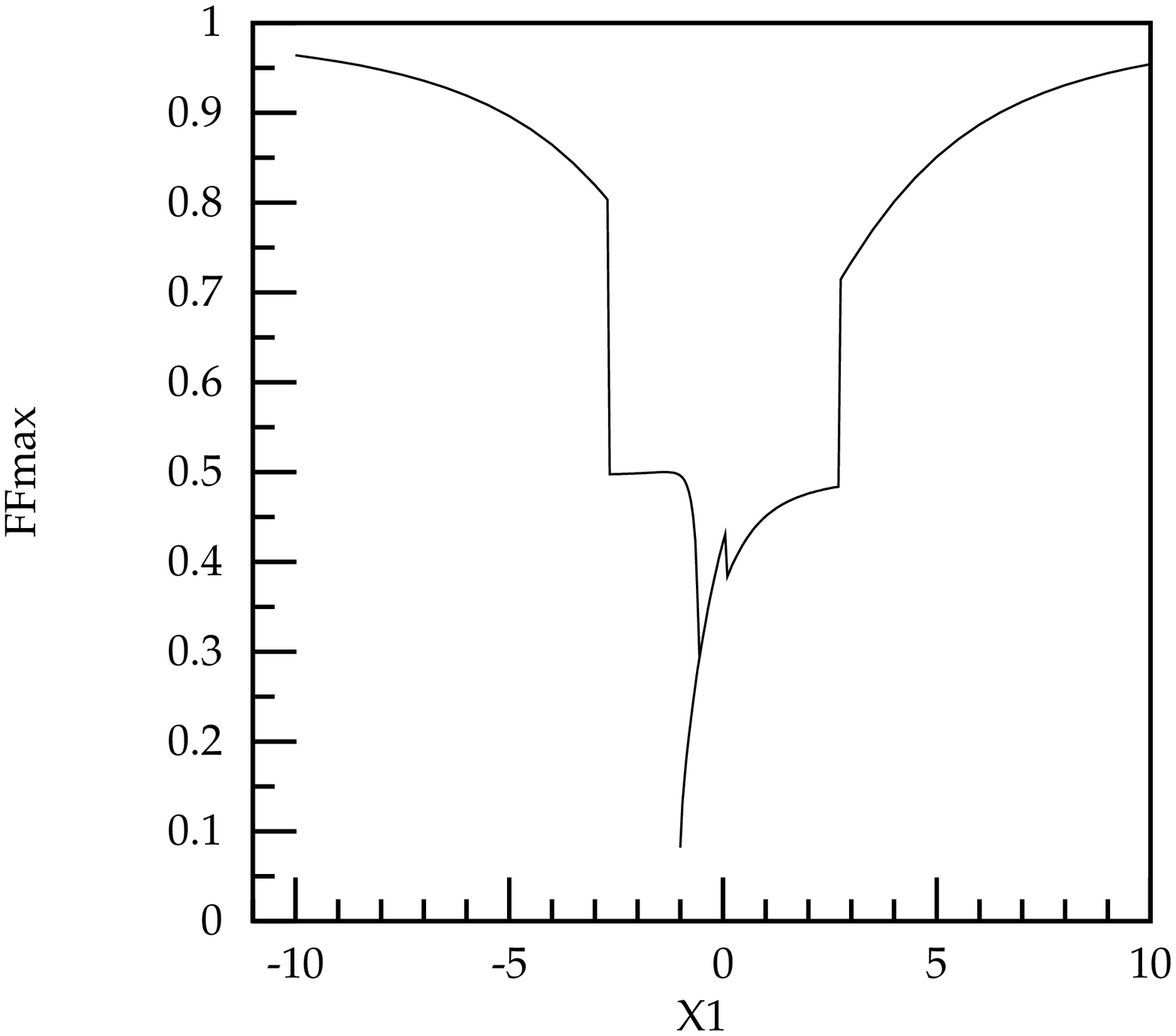}}
 \epsfxsize=8cm \put(4,0){\epsffile{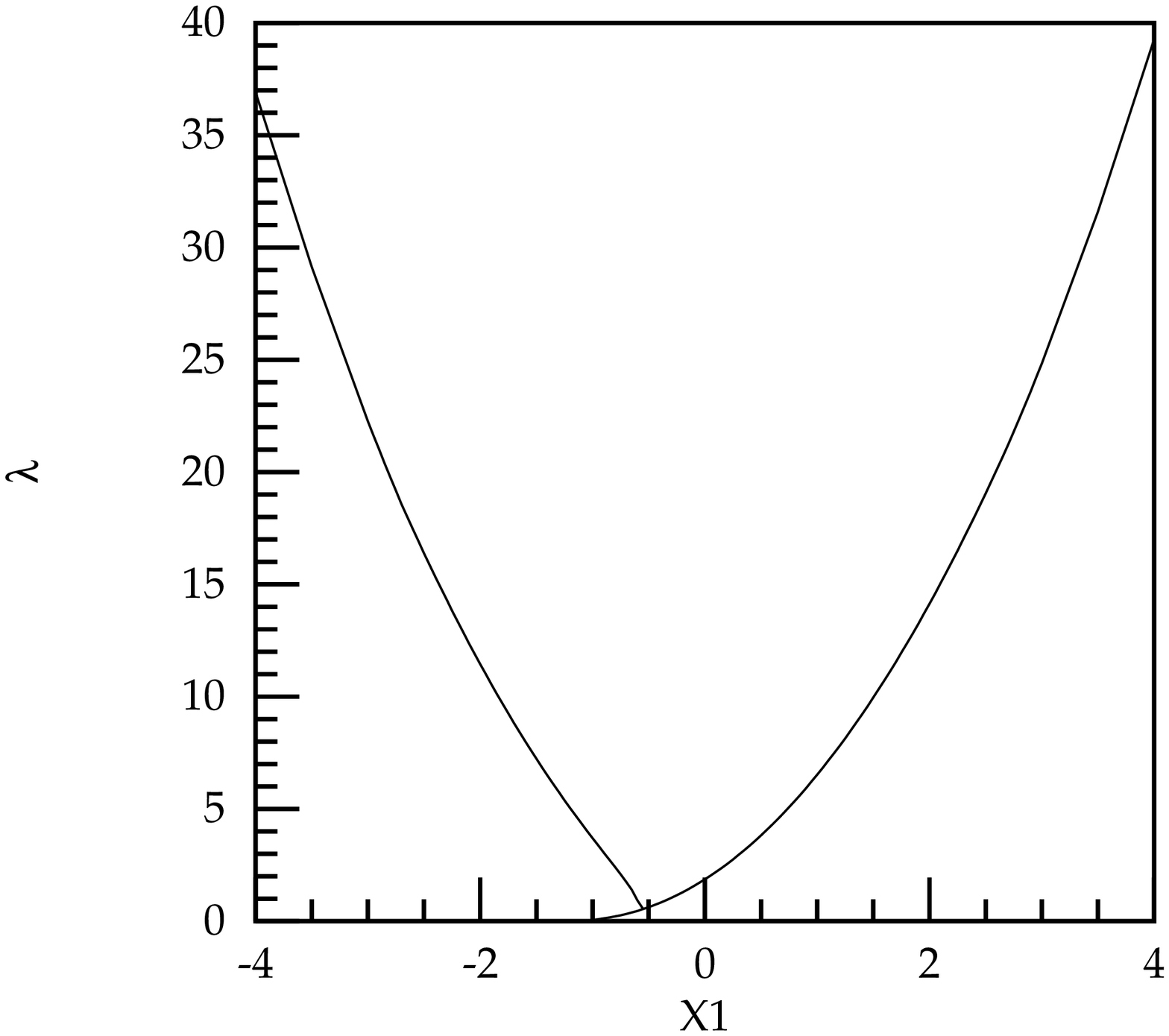}}
\end{picture}
\caption{\label{FIGsolG21.5}
Solutions of Eqs. (\ref{Eqnphi}-\ref{Eqna})
at $G_2=1.5$ and $\chi_2=0.2$. The plots of: a) $|\ph|^2$,  b) $\lambda$.}
\end{figure}

When $G_2=1.5$, the structure of the solutions is similar to that of $G_2=1$,
but it now exhibits a region where more than one solution exists (other
than the fully delocalised solution that always exists). In 
Fig. \ref{FIGsolG21.5} we plot the value of $|\ph|^2$ as well as $\lambda$
as a function of $\chi_1$ and one notices an even richer structure. Notice
that $\lambda$ corresponds to the eigenenergy with opposite sign, so a large 
value of $\lambda$ corresponds to a lower energy state.

\begin{figure}[htbp]
\unitlength1cm \hfil
\begin{picture}(12,15)
 \epsfxsize=6cm \put(0,12){\epsffile{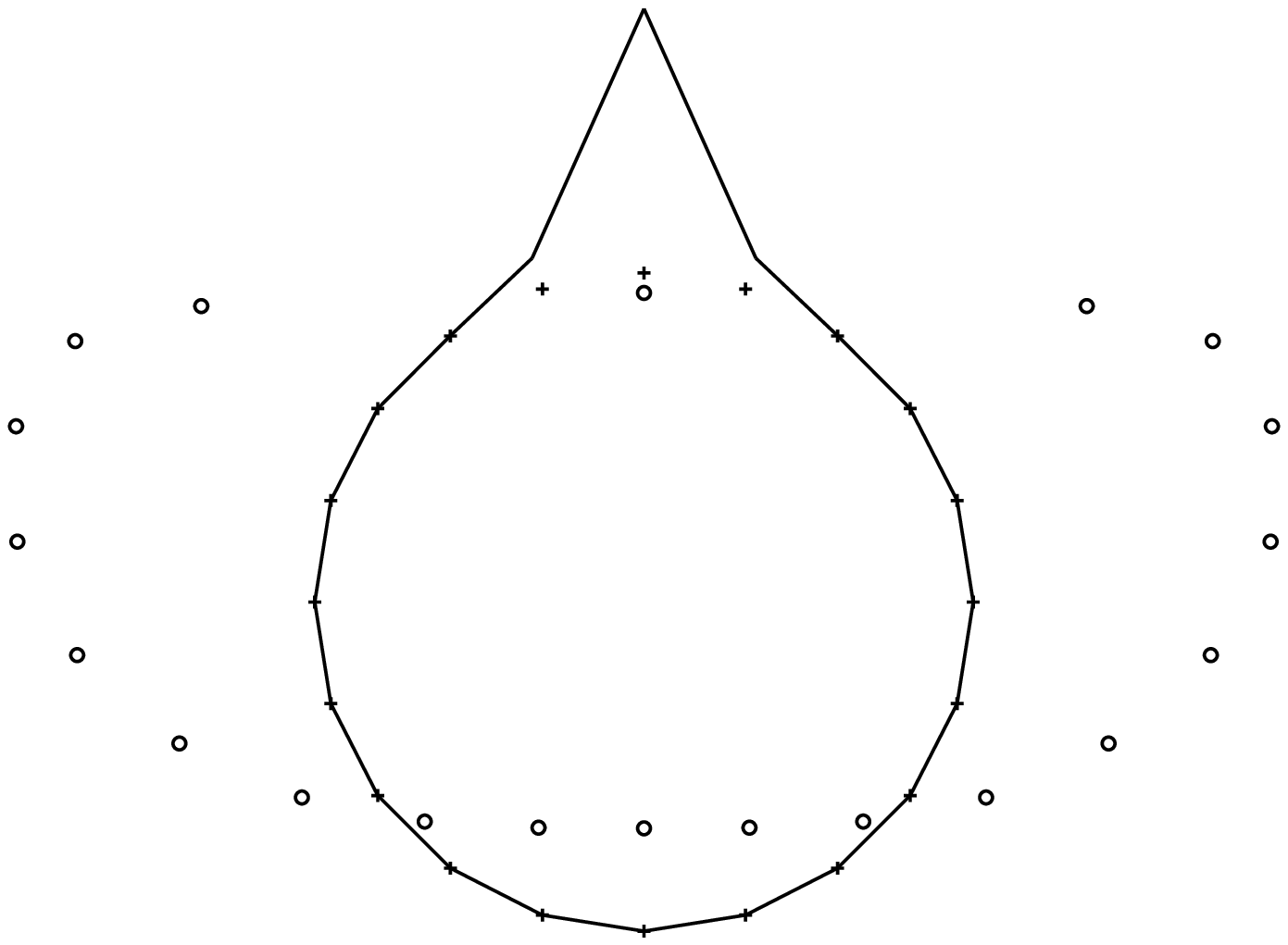}}
 \epsfxsize=6cm \put(6,12){\epsffile{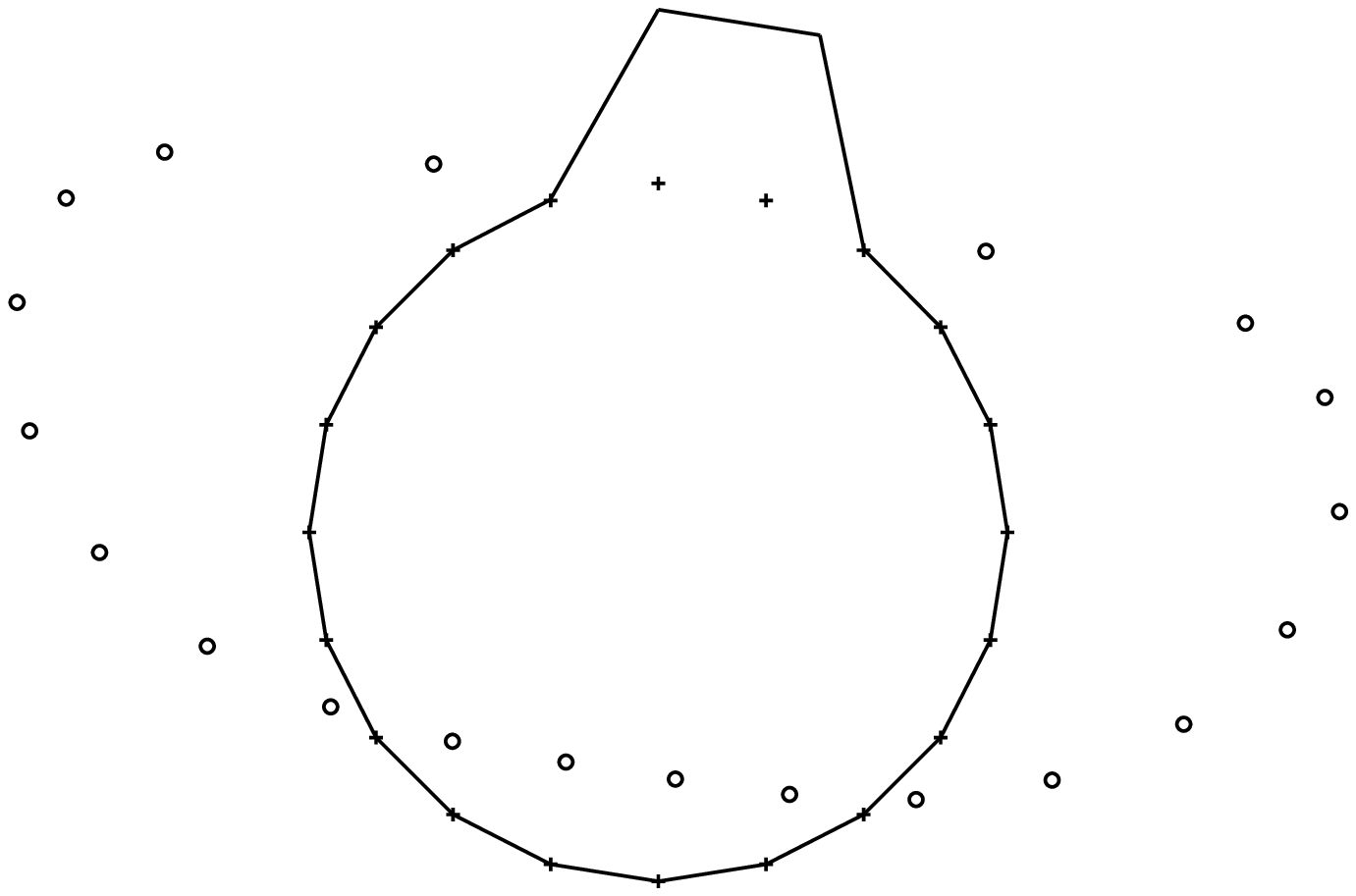}}
 \epsfxsize=6cm \put(0,8){\epsffile{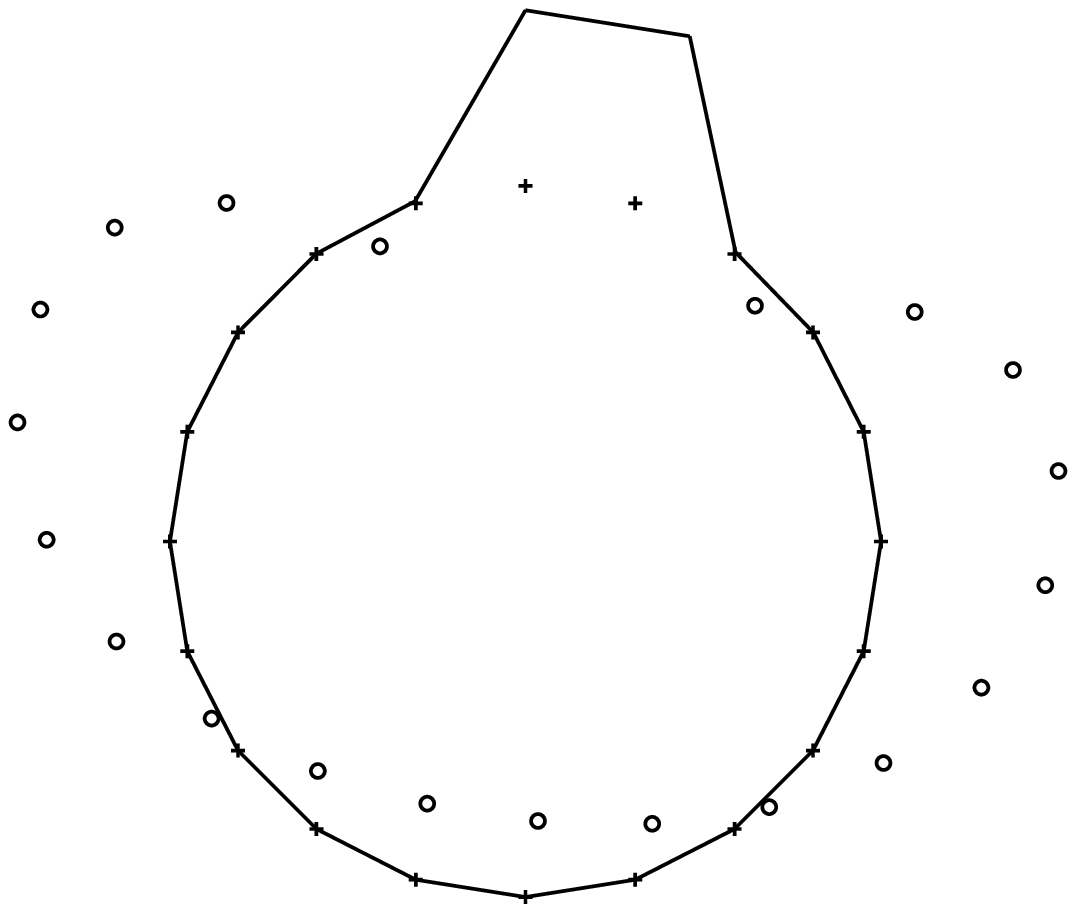}}
 \epsfxsize=6cm \put(6,8){\epsffile{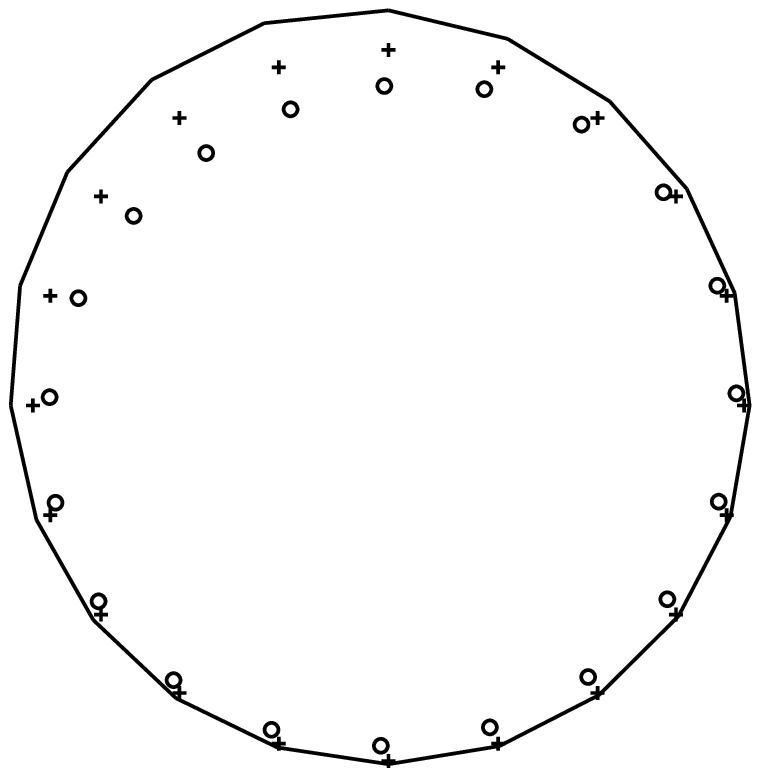}}
 \epsfxsize=6cm \put(0,4){\epsffile{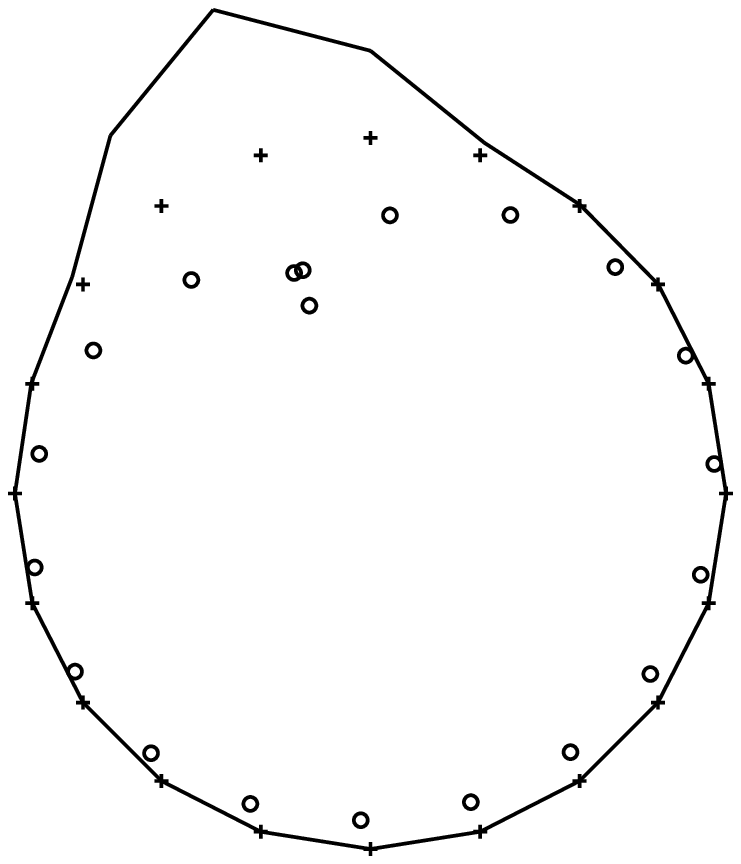}}
 \epsfxsize=6cm \put(6,4){\epsffile{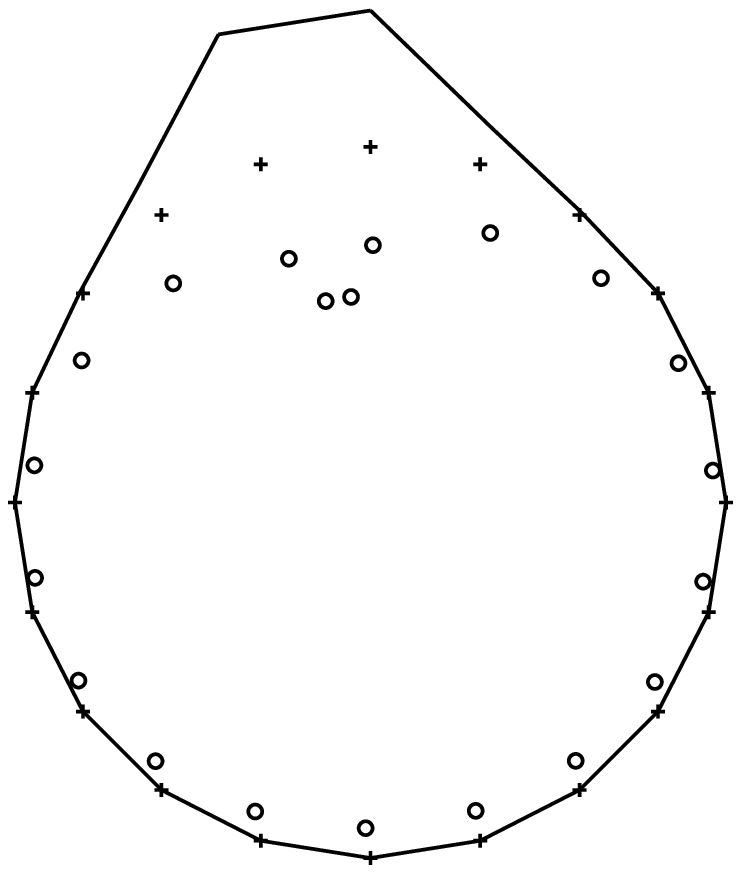}}
 \epsfxsize=6cm \put(0,-0.5){\epsffile{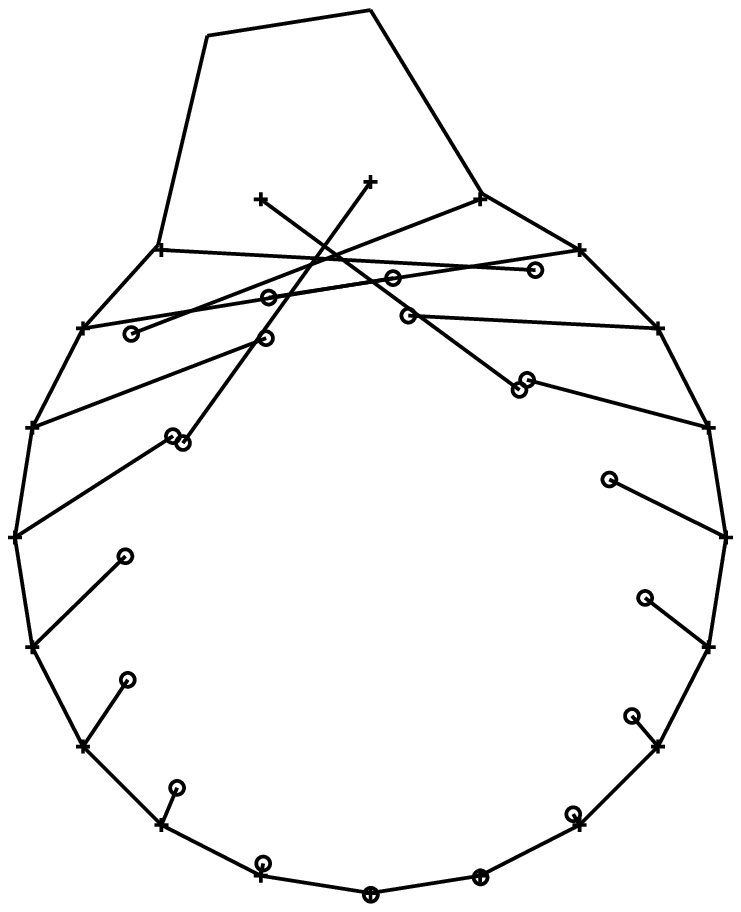}}
 \epsfxsize=6cm \put(6,-0.5){\epsffile{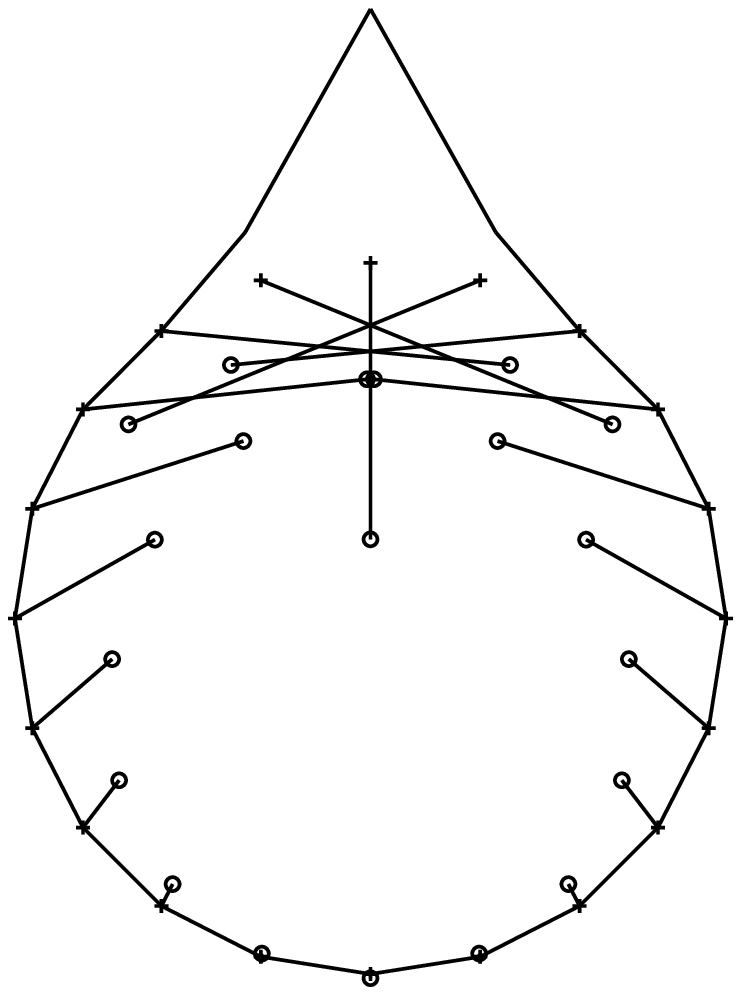}}
\put(1.5,13){a}
\put(7.5,13){b}
\put(1.5,9){c}
\put(7.5,9){d}
\put(1.5,5){e}
\put(7.5,5){f}
\put(1.5,0.5){g}
\put(7.5,0.5){h}
\end{picture}
\caption{\label{FIGsolB}
Numerical solutions of Eqs. (\ref{Eqnphi}-\ref{Eqna}) for $G_2=1.5$,
$\chi_2=0.2$ and a) $\chi_1=-2.7$ b) $\chi_1=-2.65$
c) $\chi_1=-0.95$ ($\lambda=3.38$)
d) $\chi_1=-0.95$ ($\lambda=0.09$) e) $\chi_1=0.05$
f) $\chi_1=0.1$ g) $\chi_1=2.7$ h) $\chi_1=2.75$ .
The $+$ represent the undeformed chain. 
Symbols are explained in caption of Fig. 1.
In (g) and (h), we also show where each site has been moved
to by the chain deformation.}
\end{figure}

In Fig. \ref{FIGsolB} we present several solutions for the same value of
$G_2=1.5$.
When $\chi_1=-2.7$, Fig.\ref{FIGsolB}a, $|\ph|^2 = 0.803$
and the quasiparticle is strongly localised on one site.
When $\chi_1=-2.65$, Fig. \ref{FIGsolB}b, $|\ph|^2 = 0.50$ and
the quasiparticle is strongly localised on two neighbouring sites.
When $\chi_1=-0.95$, Fig. \ref{FIGsolB}c and Fig. \ref{FIGsolB}d,
there are two solutions,  $|\ph|^2 = 0.49$ and $|\ph|^2 = 0.13$. The first one
is localised on two sites, while the second one is
delocalised and has a different symmetry. Fig. \ref{FIGsolB}b shows that
the most localised of these solutions has the lowest energy.
When $\chi_1=0.05$, $|\ph|^2 = 0.43$, Fig. \ref{FIGsolB}e,  
{\it ie} just before the little spike
in Fig. \ref{FIGsolG21.5}a, the quasiparticle is slightly delocalised
and has a symmetry axis passing through a chain site.
When $\chi_1=0.1$, $|\ph|^2 = 0.38$, Fig. \ref{FIGsolB}f, {\it ie} 
just after the little spike
in Fig. \ref{FIGsolG21.5}a, the quasiparticle is still slightly delocalised 
but it possesses a different symmetry, {\it ie} the axis of symmetry passes 
between the 2 chain sites.  
When $\chi_1=2.7$, $|\ph|^2 = 0.48$, Fig. \ref{FIGsolB}g,  
the quasiparticle is strongly localised on two sites.
When $\chi_1=2.75$, $|\ph|^2 = 0.714$, Fig. \ref{FIGsolB}h,  
the quasiparticle is localised on one site.

\begin{figure}[htbp]
\unitlength1cm \hfil
\begin{picture}(8,8)
 \epsfxsize=10cm \put(0,0){\epsffile{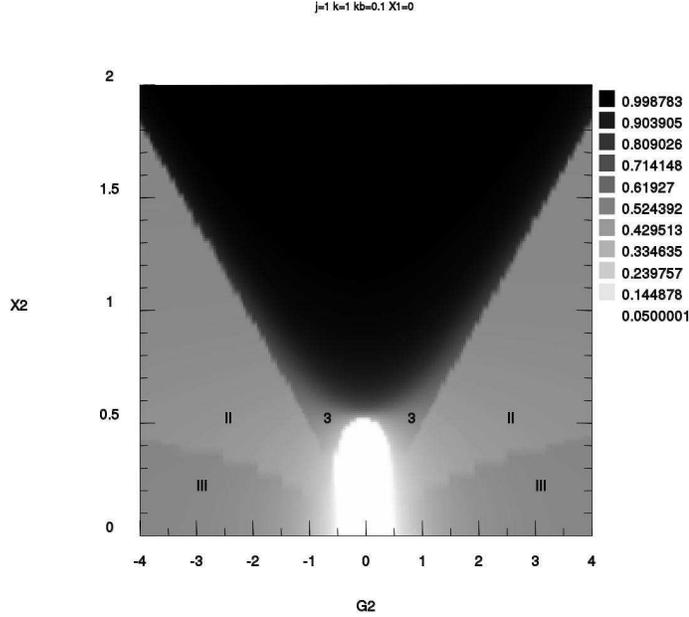}}
\end{picture}
\caption{\label{FIGplotG2X2}
Density for the solutions of Eqs. (\ref{Eqnphi}-\ref{Eqna}), $|\ph|^2$, for 
$\chi_1=0$.}
\end{figure}

\begin{figure}[htbp]
\unitlength1cm \hfil
\begin{picture}(12,15)
 \epsfxsize=6cm \put(0,12){\epsffile{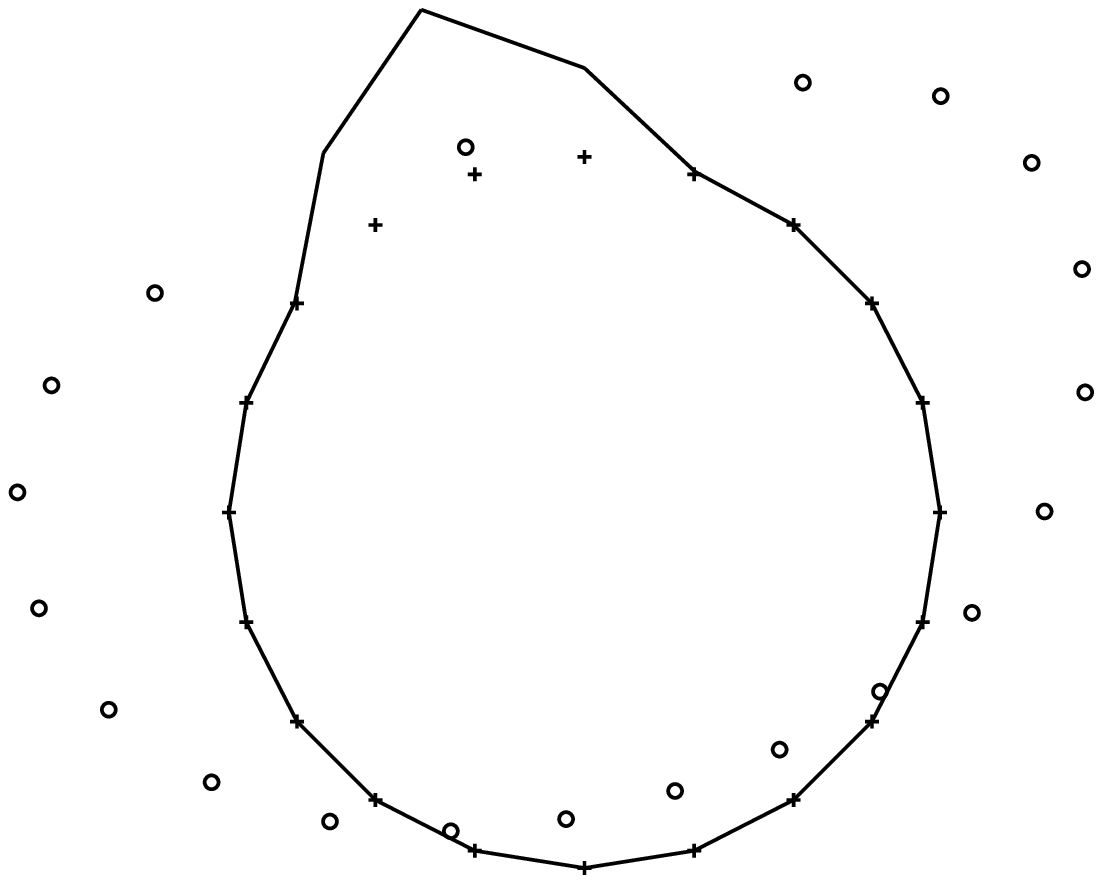}}
 \epsfxsize=6cm \put(6,12){\epsffile{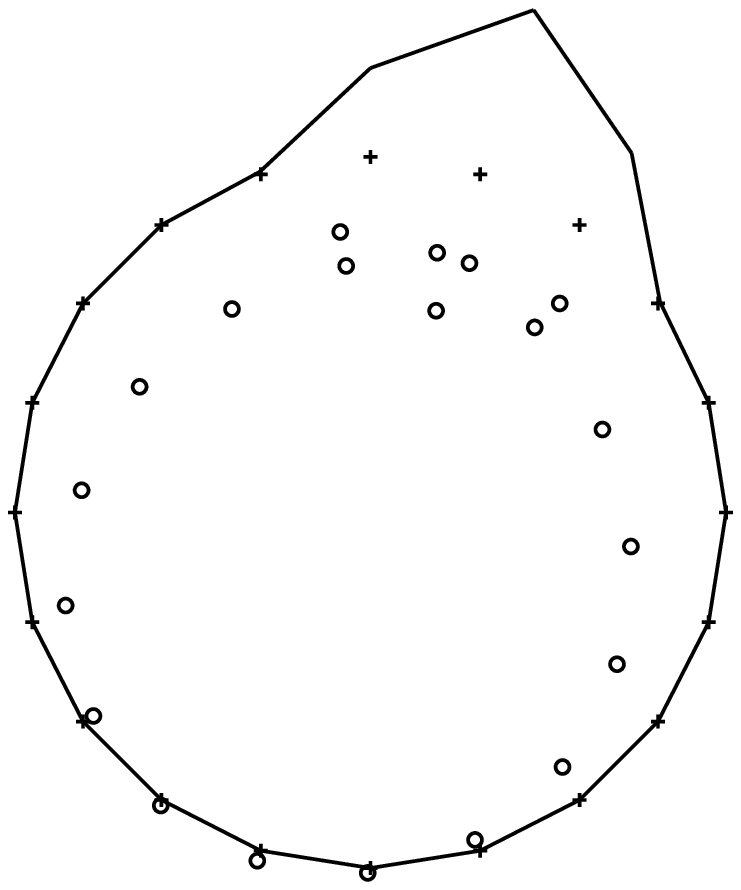}}
 \epsfxsize=6cm \put(0,8){\epsffile{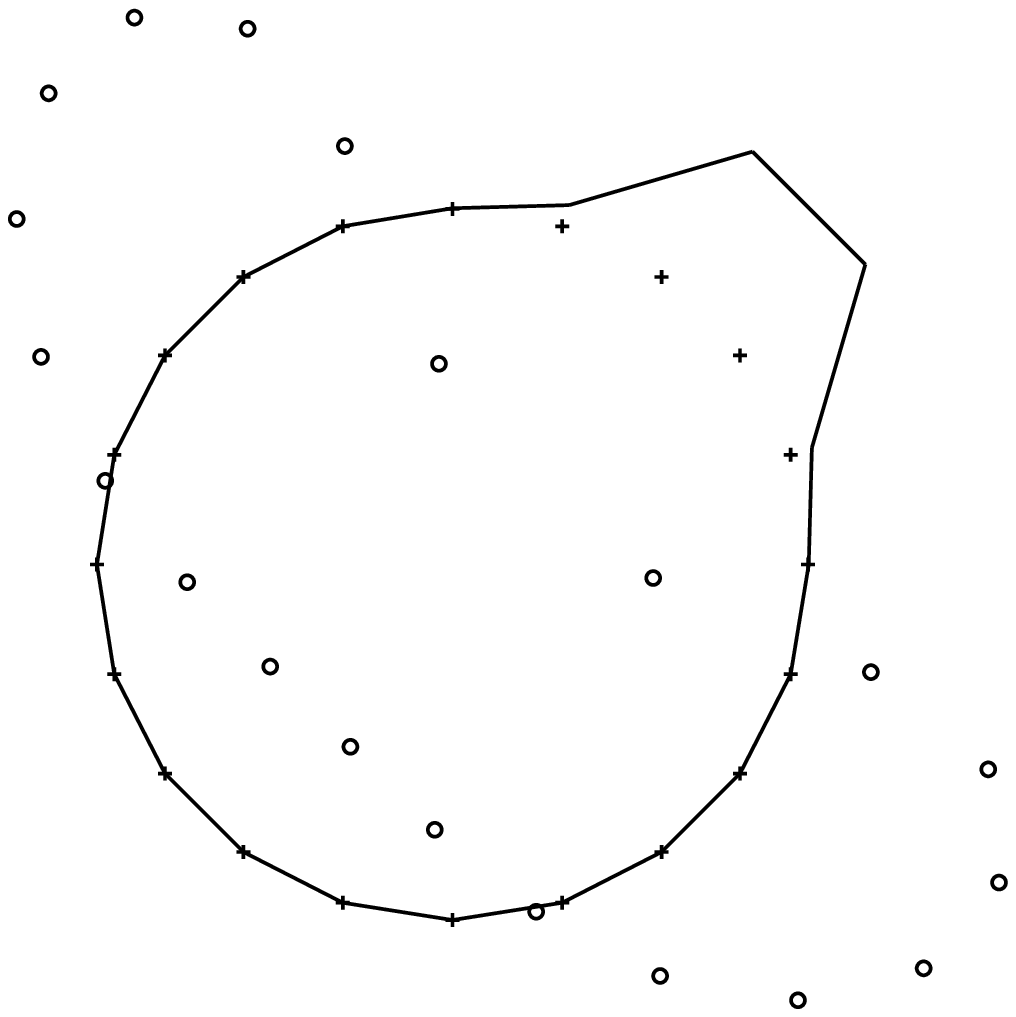}}
 \epsfxsize=6cm \put(6,8){\epsffile{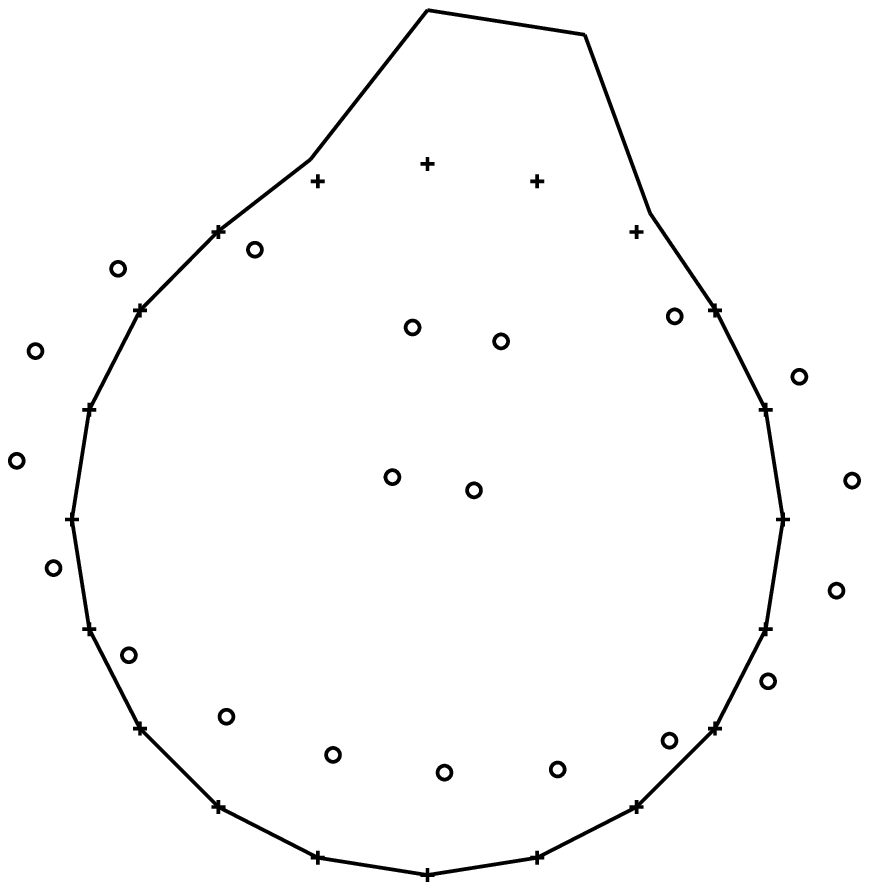}}
 \epsfxsize=6cm \put(0,4){\epsffile{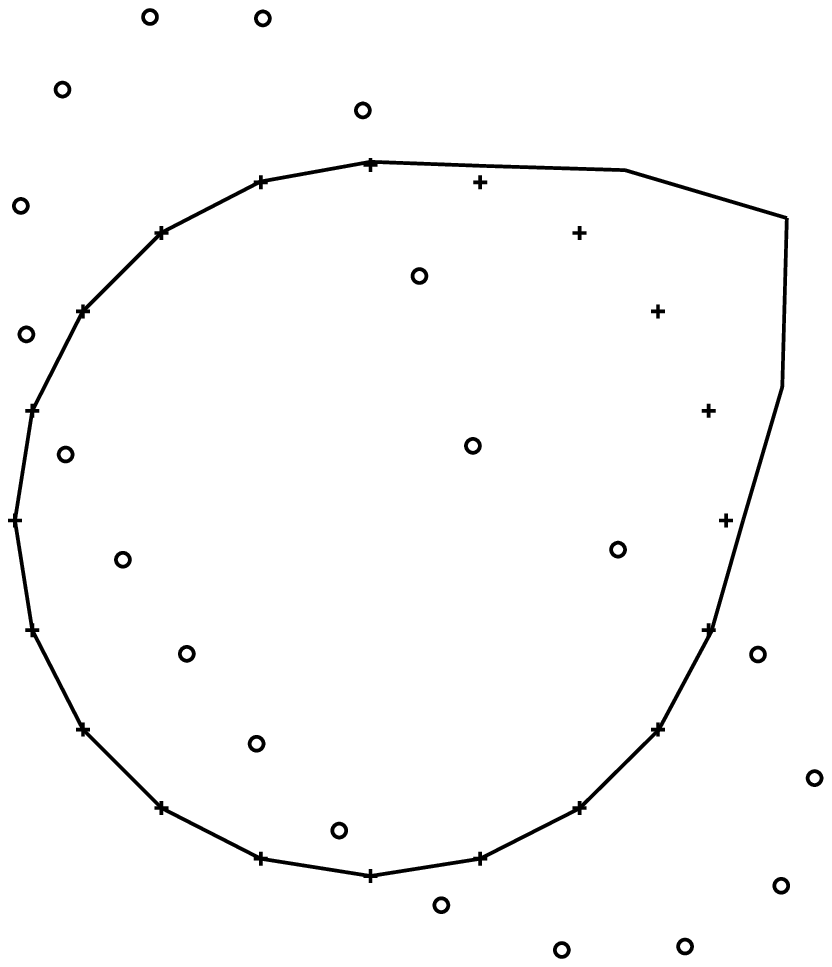}}
 \epsfxsize=6cm \put(6,4){\epsffile{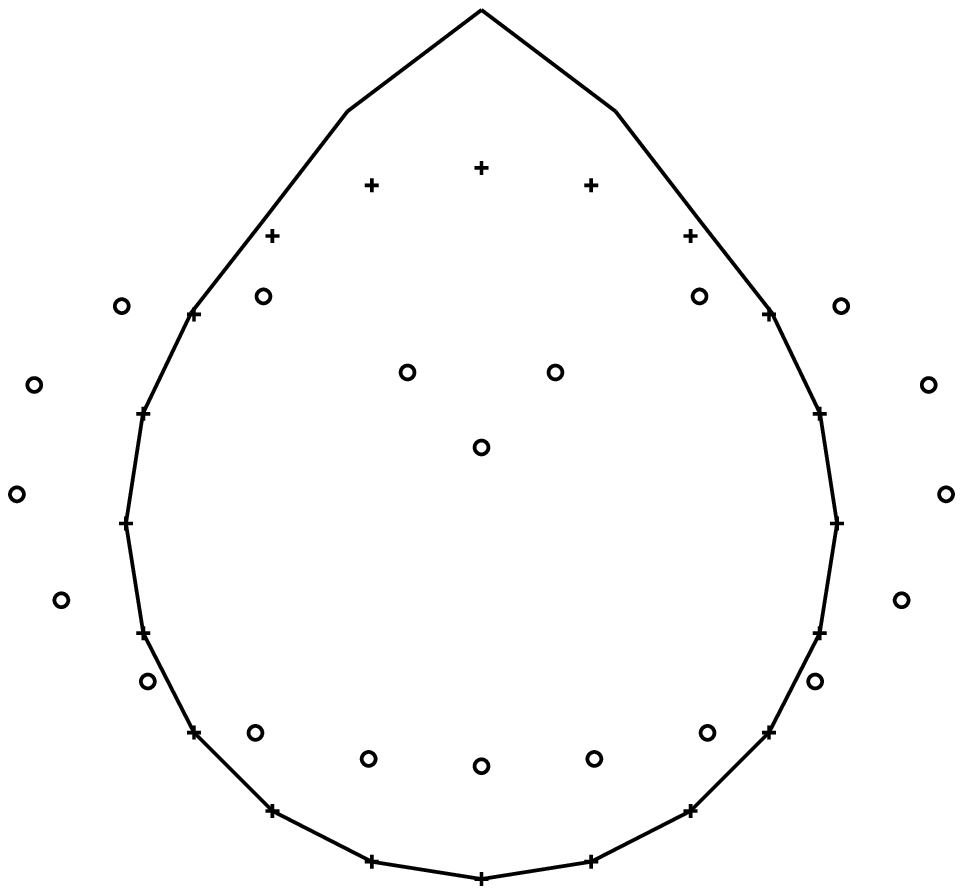}}
 \epsfxsize=6cm \put(0,-0.5){\epsffile{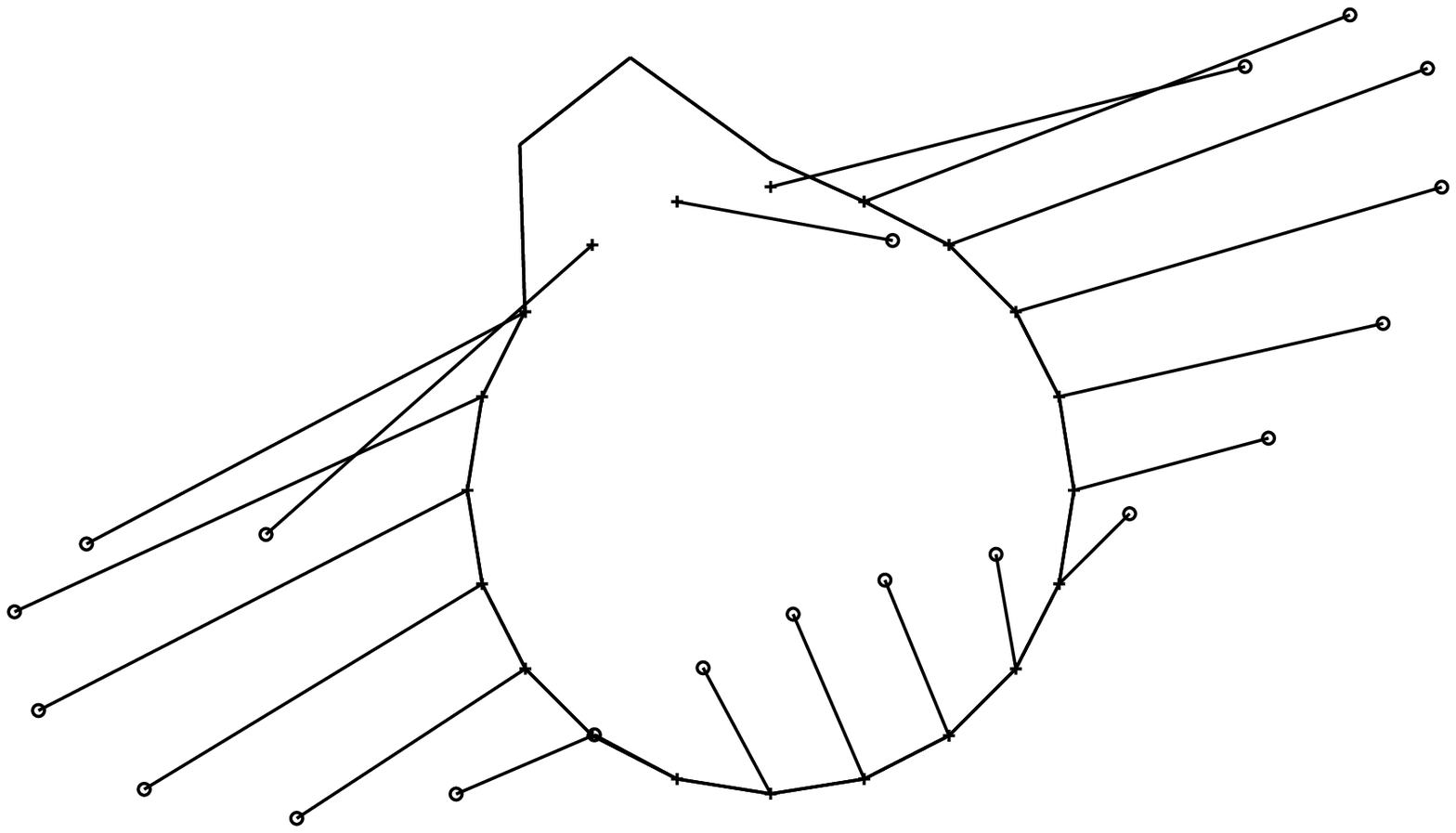}}
 \epsfxsize=6cm \put(6,-0.5){\epsffile{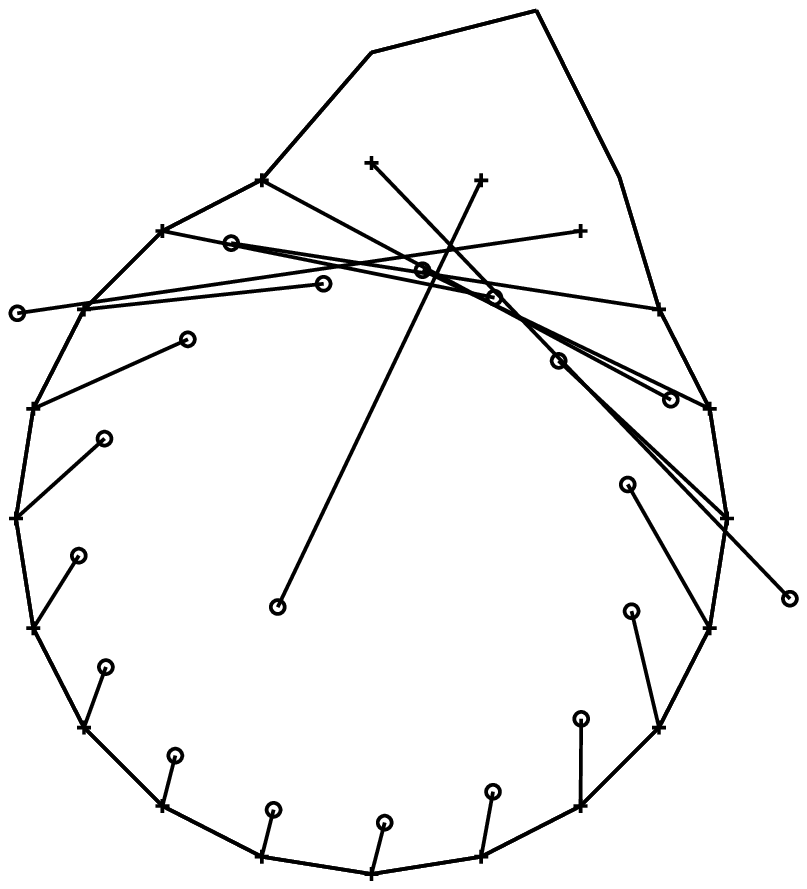}}
\put(1.5,13){a}
\put(7.5,13){b}
\put(1.5,9){c}
\put(7.5,9){d}
\put(1.5,5){e}
\put(7.5,5){f}
\put(1.5,0.5){g}
\put(7.5,0.5){h}
\end{picture}
\caption{\label{FIGsolX1}
Numerical solutions of Eqs. (\ref{Eqnphi}-\ref{Eqna}) for $\chi_1=0$ ,
a) $\chi_2=0.1$, $G_2=-3$ b) $\chi_2=0.1$, $G_2=3$
c) $\chi_2=0.5$, $G_2=-2$ d) $\chi_2=0.5$, $G_2=2$
e) $\chi_2=0.5$, $G_2=-0.7$ f) $\chi_2=0.5$, $G_2=0.7$
g) $\chi_2=0.5$, $G_2=-7$ h) $\chi_2=0.5$, $G_2=7$
Symbols are explained in caption of Fig. 1.
In (g) and (h), we also show to where each site has been moved
to by the  deformation of the chain.}
\end{figure}

In Fig. \ref{FIGplotG2X2} we present the value of $\cmod{\lambda}$ as a
function of $G_2$ and $\chi_2$ for $\chi_1=0$ while Fig. \ref{FIGsolX1}
presents some typical solutions for that range of parameters.
We notice that when $\chi_2$ and $G_2$ are small, the quasiparticle is always 
fully delocalised. When $\chi_2$ is larger than $0.5$,
there is a triangular region where the solution is strongly localised on one
lattice site. We note also that there are 3 different zones with the same
shade of grey which we have marked as III, II and 3 in Fig. \ref{FIGplotG2X2}
and which correspond to solutions spread mostly
on 3 sites (Fig. \ref{FIGsolX1}a and \ref{FIGsolX1}b),
2 sites (Fig. \ref{FIGsolX1}c and \ref{FIGsolX1}d)
and 3 sites (Fig. \ref{FIGsolX1}e and \ref{FIGsolX1}f), respectively.

The most surprising solutions, shown in Fig. \ref{FIGsolX1}g and
\ref{FIGsolX1}h, are non symmetric. They occur when $\chi_2 \ge 0.5$ and when
$G_2$ is large. When $G_2$ decreases for a fixed value of $\chi_2$,
the asymmetry of the solution decreases so that the solution becomes 
fully symmetric and the quasiparticle is localised on two sites.

Our numerical solutions show that there are various types of solutions. In
particular, they show that if
$\chi_1$, $\chi_2$ and $G_2$ are not too small, the electron-phonon 
interaction is responsible for the
localisation of the quasiparticle. In particular, a strong interaction between 
the quasiparticle and the phonon fields ($G_2$ large) introduces some
sharp transition between the different types of solutions and also leads to
an overlap between them.

It is also interesting to note that overall, the solution is either localised
on a few lattice sites or spread nearly uniformly across the entire chain,
but the range of parameters for which a broad soliton exists is very narrow.
This is shown on Fig. \ref{FIGplotG20} -\ref{FIGplotG21} by the sharp
transitions between the grey and the white zones.

\section{The Continuum Limit.}

Though the system of equations always admit localised solutions, these
solutions are not always the states of the lowest energy. At strong enough 
electron-phonon coupling the self-trapping of a
quasiparticle can take place with the formation of polaron-type states.
Depending on the strength of the coupling, the localisation can be (i) strong
with the formation of a small polaron state with the localisation of a 
quasiparticle mainly on one lattice site, or (ii) comparatively weak 
with the formation
of a large polaron state with the distribution of self-trapped quasiparticle 
among several lattice sites.

We consider this latter case here. In linear chains the solutions are
well described in the continuum approximation which is valid for large
enough systems {\it i.e.} $N \gg 1$. For a circular chain this means also 
that $\alpha$ is very small.  For such 
circles, with large enough $N$, we can neglect the term involving
$\sin(\frac{\alpha}{4})$ in (\ref{F2-F1}). In this case
Eq. (\ref{NLEqpsik}) can be rewritten in the form
\begin{eqnarray}
({\cal E}(k) &-& E)\,\psi(k)\, -
\frac{1}{N} \sum_{q,k_1}G_1(k,k_1,q) \psi^{*}(k_1) \psi(k_1+q) \psi(k-q)\,
\nonumber\\
&-&\frac{1}{N} \sum_{q,k_1(|q| > \alpha)} \frac{{\chi_2}^2}{\kappa_b}
\psi^{*}(k_1) \psi(k_1+q) \psi(k-q)\, = \,0 .
\label{NLEqpsik1}
\end{eqnarray}
In the last term on the r.h.s. of Eq. (\ref{NLEqpsik1}) we can use the
approximation $\sum_{q,k_1,|q| > \alpha}\dots = b^2 \sum_{q,k_1}\dots$.
Of course, the value of the constant $b^2\ll 1$ depends on $\alpha$ as well 
as on the degree of localisation of the  solution. For fully
delocalised solutions $b^2 = 0$ but $b^2$ rapidly tends to unity as the 
solution becomes more localised. In this way Eq. (\ref{NLEqpsik1}) becomes
\begin{equation}
({\cal E}(k) - E)\,\psi(k)\, -  \frac{1}{N} \sum_{q,k_1}
\tilde{G}(k,k_1,q) \psi^{*}(k_1) \psi(k_1+q) \psi(k-q)\,=\,0,
\label{NLEqpsik2}
\end{equation}
where
\begin{equation}
\tilde{G}(k,k_1,q) = b^2 \frac{{\chi_2}^2}{\kappa_b} + G_1(k,k_1,q).
\label{Gtilde}
\end{equation}

A weak localisation in the site representation corresponds to a strong
localisation in the $k$-representation, {\it  i.e.} $\psi(k)$ 
($k = \alpha \nu $)
is essentially nonzero only for several values of $\nu$ in the vicinity of
$\nu = 0$. In the long-wave approximation when $k = \alpha \nu \ll 1$
(which assumes $N \gg 1$) we can write down the following expansions for the
functions in (\ref{NLEqpsik}):
\begin{eqnarray}
{\cal E}(k) &=&{\cal {E}}_0\,- \,2J\,+\,Jk^2\, + \, \dots ,
\nonumber\\
G(k,k_1,q) &=& b^2\frac{{\chi_2}^2}{\kappa_b} +
4\frac{(\chi_1+G_2)^2}{\kappa }+ \dots .
 \label{approx}
\end{eqnarray}

To solve Eq.(\ref{NLEqpsik}) we introduce the function
\begin{equation}
\varphi(x)=\frac{1}{\sqrt {N}} \sum_{k} e^{ikx} \psi(k)
\label{varphi}
\end{equation}
of the continuum variable $x$. Note that $\varphi(x)$ is a periodic function,
$\varphi(x+N) = \varphi(x)$ and, at $x=n$, this is a unitary transformation
of $\psi_{\mu}(k)$ to the site representation. Moreover, if $\varphi(x)$ is
known, we can find $\psi(k)$.
\begin{equation}
\psi(k)=\frac{1}{\sqrt {N}} \int_{-\frac{N}{2}}^{\frac{N}{2}} e^{-ikx}
\varphi(x) dx .
\label{varphi2}
\end{equation}

Using the approximation (\ref{approx}), one can transform  (\ref{NLEqpsik})
into a nonlinear differential equation for $\varphi (x)$:
\begin{equation}
J\frac{d^2\varphi(x)}{dx^2}+G|\varphi (x)|^2\varphi(x)+\Lambda \varphi(x)=0,
\label{dnlse}
\end{equation}
which is the stationary nonlinear Schr{\"o}dinger equation (SNLSE).
Here $\Lambda$ is defined in (\ref{Lambda}) and
\begin{equation}
G=G(0,0,0)=4\frac{(\chi_1+G_2)^2}{\kappa }+\frac{{\chi_2}^2}{\kappa_b},
\label{G}
\end{equation}

Next we rewrite \Ref{dnlse} in the dimensionless form after defining
\begin{equation}
\Lambda=-\mu^2 J,\quad \textrm{and} \quad 
                            G = 2Jg \quad\hbox{where}\quad  g=g_l+g_b.
\quad g_b=\frac{{\chi_2}^2}{2J \kappa_b},
\label{g}
\end{equation}
where $g_l$ is given in \Ref{gl}.
We get
\begin{equation}
\frac{d^2\varphi(x)}{dx^2}+2g|\varphi (x)|^2\varphi(x)-\mu^2 \varphi(x)=0.
\label{nlse}
\end{equation}

We look for a solution of \Ref{nlse}  which satisfies the periodic boundary
condition
\begin{equation}
\varphi(x+L)=\varphi(x),
\label{p-cond}
\end{equation}
and the normalisation condition
\begin{equation}
\int _0 ^L |\varphi(x)|^2dx=1,
\label{n-cond}
\end{equation}
where $L$ is the length of the ring (in the adimensional
units used here, $L=N$). We see that $\varphi(x)$ is a real function and so we
can integrate Eq.\Ref{nlse} and obtain
\begin{equation}
\Bigl(\frac{d\varphi(x)}{dx}\Bigr)^2 =\mu^2 \varphi^2(x)-g\varphi^4 (x) + C
\label{int1}
\end{equation}
where $C$ is the integration constant. So we have to invert
\begin{equation}
x = - \int_{\varphi(0)}^{\varphi(x)}\frac{d\varphi}
{\sqrt{\mu^2 \varphi^2-g\varphi^4 + C}} .
\label{int2}
\end{equation}

Using the substitution
\begin{equation}
\varphi^2 = \frac{1}{g}(\frac{\mu^2}{3} - \tau), 
\label{subst}
\end{equation}
we reduce the problem \Ref{int2} to a normal Weierstrass form
\begin{equation}
x = \int_{\mathcal{F}(0)}^{\mathcal{F}(x)}\frac{d\tau}
{\sqrt{4 \tau^3-g_2\tau - g_3}}
\label{intWf}
\end{equation}
which is solved by the Weierstrass elliptic function $\wp (x + \it{c})$
\cite{Akhiez, BatErd} ($\it{c}$ being some constant). Here, and
in what follows, we shall use the standard notations for the theory of elliptic
functions. In our case, the constants $g_2$ and $g_3$ in \Ref{intWf}
are determined by
\begin{eqnarray}
g_2 &=& 4(\frac{1}{3}\mu^4 + gC), \nonumber\\
g_3 &=& - \frac{4}{3}\mu^2 (\frac{2}{9}\mu^4 + gC).
  \label{g23}
\end{eqnarray}

The elliptic Weierstrass function $\wp (z;g_2,g_3) =\wp (z|\omega,\omega')$
is a two-parametric doubly periodic function in the complex plane $z=x+iy$. 
The two parameters are either the constants $g_2$ and $g_3$ or, equivalently, 
the two periods, which are denoted as $2\omega$ and $2\omega'$, 
of the function 
$\wp (z + 2m\omega + 2n\omega') =\wp (z)$ with $m$ and $n$ being integers. 
In the theory of elliptic functions \cite{Akhiez, BatErd} the following 
symmetric notations are generally used
\begin{equation}
\omega _1=\omega ,\quad \omega _2 = -\omega  -\omega' ,\quad
\omega _3 = \omega ' 
\end{equation}
so that $\wp (\omega _j) = \it{e}_j$ ($j=1,2,3$) where $\it{e}_j$
are the roots of the cubic equation:
\begin{equation}
4 \tau^3-g_2\tau - g_3 = 0 .
\label{cubeq}
\end{equation}

By definition, $|\varphi|^2 $ is a real and bounded function of the real
coordinate $x$. Taking into account the analytical properties of the
Weierstrass function, we conclude that $\varphi$ defined by 
\Ref{intWf} will be physically meaningful only when the discriminant
$g_2^3 - 27g_3^2$ of the cubic equation \Ref{cubeq} is positive and all three
roots $\it{e}_j$ are real and different. In the theory of elliptic
functions \cite{BatErd} the real roots $\it{e}_j$ of the cubic equation
\Ref{cubeq} are distributed as follows:
\begin{equation}
\it{e}_1 > \it{e}_2 > \it{e}_3 \quad \mbox{\rm and} \quad
\it{e}_1 >0, \quad \it{e}_3 <0 .
\label{rootdistr}
\end{equation}
In our case $g_2^3 - 27g_3^2 =
16 g^2C^2(\mu^4 + 4gC)$ and the roots of the cubic equation \Ref{cubeq} are
\begin{equation}
\frac{1}{3}\mu^2, \qquad
\frac{1}{2}(-\frac{1}{3}\mu^2 \pm \sqrt{\mu^4 + 4gC}) ,
\label{roots}
\end{equation}
{\it i.e.} for the integration constant $C$ we have the constraint 
$4gC \geq -\mu^2$.
In this case the period
$2\omega$ is real,  $2\omega'$ is purely imaginary, and the solution of
\Ref{intWf} reads as $\wp (x + \omega')$. Therefore the solution of
Eq.\Ref{int1} (or \Ref{int2}) is given by
\begin{equation}
\varphi^2 (x) = \frac{1}{g}(\frac{\mu^2}{3} - \wp (x + \omega')) .
\label{solut}
\end{equation}
This solution must satisfy the boundary \Ref{p-cond} and normalisation
\Ref{n-cond} conditions which lead to the relations
\begin{eqnarray}
2 m \omega &=& N, \nonumber\\
\frac{1}{3}\mu^2 N + 2m \eta &=& g.
  \label{relW}
\end{eqnarray}
Here $m=1,2,\dots$ is an arbitrary number and $\eta = \zeta (\omega)$ is the
constant of the elliptic functions theory ($\zeta (z)$ is the Weierstrass
$\zeta$-function determined by the relation $\wp (z) = -\zeta '(z)$).
The conditions \Ref{relW} allow us to determine the integration constant $C$
and the eigenvalue $\mu^2$, and then to calculate the total energy of the 
system which includes the energy of the deformation. Using solution
\Ref{solut} we get the following expression for the total energy:
\begin{equation}
E_{tot} = {\cal E}(0) - \epsilon J , \quad
\epsilon = \frac{1}{3} \mu^2 \,-\,\frac{1}{3}CN.
\label{Etot}
\end{equation}

To solve the system of equations \Ref{relW}, let us introduce the parameter
\begin{equation}
{\it{k}}^2 =
\frac{\it{e}_2-\it{e}_3}{\it{e}_1-\it{e}_3} \leq 1
\label{modulk}
\end{equation}
which, from now on, will be considered as an independent parameter instead of 
the integration
constant $C$. Here $\it{e}_j = \wp (\omega _j)$ are the roots the cubic
equation \Ref{cubeq} and $\it{k}$ is the modulus of the  elliptic Jacobi
functions \cite{BatErd}. Using the formulas
\begin{equation}
\bf{K}(\it{k}) = (\it{e}_1 - \it{e}_3)^{\frac{1}{2}} \omega ,\qquad
\bf{E}(\it{k}) = \frac{\it{e}_1 \omega + \eta}
{(\it{e}_1 - \it{e}_3)^{\frac{1}{2}}},
\end{equation}
where $\bf{K}(\it{k})$ and $\bf{E}(\it{k})$ are complete elliptic integrals
of the first and second kind, respectively, we can rewrite
the conditions \Ref{relW} in the form
\begin{eqnarray}
2 m \bf{K}(\it{k}) = N (\it{e}_1 - \it{e}_3)^{\frac{1}{2}}&,& 
\nonumber\\
2m (\it{e}_1 - \it{e}_3)^{\frac{1}{2}} \bf{E}(\it{k}) -
(\it{e}_1 - \frac{1}{3}\mu^2 ) N  &=& g.
  \label{relJac}
\end{eqnarray}
Inserting the first expression into the second one, we can rewrite 
\Ref{relJac} as
\begin{eqnarray}
2 m \bf{K}(\it{k}) &=& N (\it{e}_1 - \it{e}_3)^{\frac{1}{2}},
\nonumber\\
(\it{e}_1 - \it{e}_3) \bf{E}(\it{k}) -
(\it{e}_1 - \frac{1}{3}\mu^2 ) \bf{K}(\it{k})
&=& \frac{g}{2m} (\it{e}_1 - \it{e}_3)^{\frac{1}{2}}.
  \label{relJac2}
\end{eqnarray}

The solution \Ref{solut} can be rewritten in the form
\begin{equation}
\varphi^2 (x) = \frac{1}{g}\Bigl( (\frac{\mu^2}{3} -\it{e}_3) -
\frac{(\it{e}_1 - \it{e}_3)(\it{e}_2 - \it{e}_3)}{\wp (x) - \it{e}_3}\Bigr) .
\label{solutb}
\end{equation}

Depending on the sign of the integration constant $C$, the roots \Ref{roots} 
will be enumerated in two different ways according to the rule \Ref{rootdistr}:
\begin{eqnarray}
\it{e}_1 &=& \frac{1}{2}(\sqrt{\mu^4 + 4gC}-\frac{1}{3}\mu^2 ),
\nonumber\\
\it{e}_2 &=& \frac{1}{3}\mu^2, \nonumber\\
\it{e}_3 &=& - \frac{1}{2}(\sqrt{\mu^4 + 4gC}+\frac{1}{3}\mu^2 )
  \label{Cpositive}
\end{eqnarray}
at $C>0$ and
\begin{eqnarray}
\it{e}_1 &=& \frac{1}{3}\mu^2, \nonumber\\
\it{e}_2 &=& \frac{1}{2}(\sqrt{\mu^4 + 4gC}-\frac{1}{3}\mu^2 ),
 \nonumber\\
\it{e}_3 &=& - \frac{1}{2}(\sqrt{\mu^4 + 4gC}+\frac{1}{3}\mu^2 )
  \label{Cnegaitive}
\end{eqnarray}
at $C<0$.
Two different solutions correspond to these two cases of root distribution.

1) $C>0$. In this case, taking into account the root distribution
\Ref{Cpositive}, we obtain from \Ref{modulk} the following expression
for the constant $C$ in terms of $\it{k}$:
\begin{equation}
C = \frac{\it{k}^2 (1-\it{k}^2)}{(2\it{k}^2-1)^2} \frac{\mu^4}{g} .
\label{C_1}
\end{equation}
The second condition in \Ref{relJac2} gives us the expression for $\mu$
\begin{equation}
\mu = \frac{g}{2m} \frac{\sqrt{2\it{k}^2-1}}{\bf{E}(\it{k}) -
(1 - \it{k}^2 ) \bf{K}(\it{k})}
\label{mu_1}
\end{equation}
and the first one gives us the equation which determines the constant $\it{k}$,
the values of which are restricted by the condition that $\it{k}>\frac{1}{2}$, 
namely:
\begin{equation}
 \bf{K}(\it{k}) \Bigl(\bf{E}(\it{k}) - (1 - \it{k}^2 ) \bf{K}(\it{k})\Bigr)
= \frac{gN}{(2m)^2}
\label{mu_2}
\end{equation}

Using the expression for the elliptic Jacobi functions in terms of the 
Weierstrass function,
\begin{equation}
cn^2 (u,\it{k})=\frac{\wp (x) - \it{e}_1}{\wp (x) - \it{e}_3}, \quad
u = (\it{e}_1 - \it{e}_3)^{1/2} x
\end{equation}
we find that the solution \Ref{solutb} is given by
\begin{equation}
\varphi(x)=\frac{\sqrt{g}\it{k}}{2m[\bf{E}-(1-\it{k}^2)\bf{K}]}
cn \left[\frac{2m\bf{K}x}{N},\it{k}  \right],
\label{cn-sol}
\end{equation}
From the periodic properties of the function $cn(u,\it{k})$,
$cn(u+4\bf{K},\it{k})=cn(u,\it{k})$,
it follows that the periodic condition \Ref{p-cond} will be satisfied
if $m=2n$, $n=1,2,...$. The total energy \Ref{Etot} of the nanocircle in
these states is then 
\begin{equation}
E_{tot}(n) = {\cal E}(0)\,-\,\frac{Jg^2}{48 n^2}
\frac{(2\it{k}^2-1)\bf{E}-(3\it{k}^2-1)(1-\it{k}^2)\bf{K}}
{[\bf{E}-(1-\it{k}^2)\bf{K}]^3}.
\label{Etotcn}
\end{equation}

2) $C<0$. In this case the roots are distributed as \Ref{Cnegaitive} and
from \Ref{modulk} it follows that
\begin{equation}
C = - \frac{1-\it{k}^2}{(2 - \it{k}^2)^2} \frac{\mu^4}{g} .
\label{C_2}
\end{equation}
The second condition in \Ref{relJac2} gives us the expression for $\mu$
\begin{equation}
\mu = \frac{g}{2m} \frac{\sqrt{2-\it{k}^2}}{\bf{E}(\it{k})}
\label{mu2}
\end{equation}
and the first one gives the equation for determining the constant $\it{k}$
\begin{equation}
\bf{E}(\it{k})\bf{K}(\it{k}) = \frac{gN}{(2m)^2},
\label{leng}
\end{equation}
Taking into account the relation
\begin{equation}
dn^2 (u,\it{k})=\frac{\wp (x) - \it{e}_2}{\wp (x) - \it{e}_3}, 
\end{equation}
we find that in this case a normalised periodic solution can be expressed
in terms of the elliptic Jacobi function as follows:
\begin{equation}
\varphi(x)=\frac{\sqrt{g}}{2m\bf{E}(\it{k})}
dn \left[\frac{2m\bf{K}(\it{k})x}{N},\it{k}  \right].
\label{dn-sol}
\end{equation}
Using the periodic properties of the function $dn(u,\it{k})$,
$dn(u+2\bf{K},\it{k})=dn(u,\it{k})$,
it follows that the number $m$ can take the values $m=1,2,...$.
The total energy \Ref{Etot} of the nanocircle in these states is then
\begin{equation}
E_{tot}(m) = {\cal E}(0)\,-\,
\frac{Jg^2}{12m^2}\frac{\bf{E}(\it{k})(2-\it{k}^2) +
\bf{K}(\it{k})(1-\it{k}^2)}{\bf{E}^3(\it{k})}.
\label{etot2m}
\end{equation}

The analysis of Eqs. \Ref{Etotcn} and \Ref{etot2m} shows that the energy of
states \Ref{dn-sol} is always lower than the energy of states \Ref{cn-sol}. 
Because we are interested in the solutions which correspond to the lowest 
energy we can ignore the
solution \Ref{cn-sol} and consider only the solution
\Ref{dn-sol}. Moreover, we are interested  only in the
ground state of the circle which corresponds to the value $m=1$ in 
\Ref{dn-sol}, \Ref{etot2m} and \Ref{leng}. All other states are excited 
states of the circle.

From \Ref{mu2} and \Ref{etot2m} we obtain the following
expressions for the eigenvalue
\begin{equation}
\lambda=-\mu^2= -\frac{g^2}{4} \frac{(2-k^2)}{E^2(k)} ,
\label{eigen2}
\end{equation}
and for the total energy
\begin{equation}
\epsilon=\frac{g^2}{12}\frac{E(k)(2-k^2)+K(k)(1-k^2)}{E^3(k)}
\label{etot2}
\end{equation}
of the quasiparticle ground state in the chain.

Let us consider next the case when the modulus of the elliptic function is
small, $\it{k}^2 \ll1$, which corresponds to the case of small nonlinearities
and not too long chains. In this case, using the expansions of complete
elliptic integrals in series of $\it{k}$,
\begin{eqnarray}
\bf{K}(\it{k}) = \frac{\pi}{2}\left(1+\frac{1}{4}\it{k}^2 +
\frac{3^2}{8^2}\it{k}^4 - \dots \right), \nonumber\\
\bf{E}(\it{k}) = \frac{\pi}{2}\left(1-\frac{1}{4}\it{k}^2 -
\frac{3}{8^2}\it{k}^4 - \dots \right)
 \label{e2}
\end{eqnarray}
we have from \Ref{leng}, at $m=1$ that
\begin{equation}
\it{k}^2=4 \sqrt{2}\sqrt{\frac{gL}{\pi^2}-1}.
\label{k2}
\end{equation}
So we can conclude that the solution exists only when $g$ exceeds
the critical value of the nonlinearity constant
\begin{equation}
g_{cr}(N)=\frac{\pi^2}{N}
\label{gcr}
\end{equation}
which depends on the length of the ring.

In this case the solution \Ref{dn-sol} can be written as
\begin{equation}
\varphi(x) \approx \frac{\sqrt{g}}{\pi \left(1-\sqrt{\frac{2gL}{\pi^2}-2}
\right)} \left[1-2 \sqrt{\frac{2gL}{\pi^2}-2} \sin ^2\frac{gx}{\pi
\left(1-\sqrt{\frac{2gL}{\pi^2}-2}\right)}\right]
\label{phi2}
\end{equation}
For the eigenvalue and the total energy at $\it{k}^2 \ll 1$ expressions
\Ref{eigen2} and \Ref{etot2} give
\begin{equation}
\lambda _{dn} \approx -\frac{2 g^2}{\pi ^2} \left(1+
\frac{1}{32}\it{k}^4 \right), \quad
\epsilon_{dn} \approx \frac{g^2}{\pi ^2} \left(1-
\frac{1}{32}\it{k}^4 \right),
\label{eigen3}
\end{equation}
Remember, that this solution exists only when $g$ exceeds the critical value
$g_{cr}$ \Ref{gcr}. Below $g_{cr}$ the fully delocalised state is realised
on the circle. At $g \rightarrow g_{cr}$ the solution \Ref{phi2}
formally tends to a delocalised solution but comparing \Ref{eidel} with
\Ref{eigen3} shows that even at $g=g_{cr}$ the total energy of the
state \Ref{dn-sol} is lower than energy of the fully delocalised state. Only in
the case $g_b=0$ does the energy of the fully delocalised solution coincide 
with
the energy of the periodic solution at $g=g_{cr}$. At $g>g_{cr}$ the energy
of the periodic solution is always lower than the energy of the fully
delocalised state.

Let us consider now the opposite case, {\it i.e.} when the modulus of the 
elliptic
function is close to 1, $\it{k}^2 \approx 1$, $\it{k}_1^2=1-\it{k}^2 \ll 1$.
This case, as we will see below, corresponds to  long chains and not too
small  nonlinearity parameters, such that the condition $gN \gg 1$ is
satisfied. Then, using the expansion of the elliptic functions and
integrals in power series of $\it{k}_1^2$,
\begin{eqnarray}
\bf{K}(\it{k}) = \ln\frac{4}{\it{k}_1} + \frac{1}{4}
\left(\ln\frac{4}{\it{k}_1} - 1 \right) \it{k}_1^2 + \dots ,  \nonumber\\
\bf{E}(\it{k}) = 1 + \frac{1}{2} \left(\ln\frac{4}{\it{k}_1} -
\frac{1}{2} \right) \it{k}_1^2 + \dots ,
 \label{e1}
\end{eqnarray}
we get from \Ref{leng}:
\begin{equation}
\it{k}_1=4 e^{-gN/4},\qquad \it{k}^2=1-16 e^{-gN/2}.
\label{k1-q}
\end{equation}
Substituting this into \Ref{dn-sol}, we get
\begin{eqnarray}
\varphi(x) \approx \frac{\sqrt{g}}{2[1+2(gN-2)e^{-gN/2}]}
\left[ 1 + \right. \nonumber \\
\left.+e^{-gN/4}\left(\frac{gx}{2}+\sinh
\frac{gx}{2} \cosh \frac{gx}{2} \right)\tanh \frac{gx}{2}
\right]\cosh ^{-1} \frac{gx}{2},
\label{dn1}
\end{eqnarray}
For the eigenenergy and the total energy we get from \Ref{eigen2} and
\Ref{etot2}
\begin{equation}
\lambda _{dn} \approx -\frac{g^2}{4} \left(1-
8(2gN-3) e^{-gN/2} \right), \quad
\epsilon_{dn} \approx \frac{g^2}{12} \left(1+
24 e^{-gN/2} \right).
\label{eigen4}
\end{equation}

We see that at large enough $g$ ($gN \gg 1$) the solution \Ref{cn-sol}
describes a well localised state and that increasing $g$ leads to the
increase  of the localisation. But it is necessary to keep in mind that the
applicability of the continuum approximation is restricted by the condition 
that the localisation is not too strong which itself imposes the 
constraint $\mu < \pi/2$.

Using the obtained solution it is possible to calculate the configuration of 
the
circle in the state \Ref{cn-sol}. Taking into account \Ref{transf1},
\Ref{transf2}, \Ref{Qjq}, and the explicit expressions for the corresponding
functions \Ref{Omegaj2}, \Ref{Gs}, \Ref{phy-varphi}, \Ref{omegas-f} and
\Ref{thetaq} we can write 
\begin{equation}
u_n\ =\ \frac{1}{N}\sum_{q}e^{iqn}U(q), \qquad
s_n\ =\ \frac{1}{N}\sum_{q}e^{iqn}S(q),
\label{displ}
\end{equation}
with
\begin{eqnarray}
U(q)\ &=&\ i\ \sum_{k} \frac{
\frac{\chi_2}{4\kappa_b} \sin(\frac{\alpha}{2}) \cos(\frac{q}{2}) +
\cos(\frac{\alpha}{2}) \sin^2(\frac{q}{2})\bigl[
\frac{\chi_1}{\kappa}\cos(\frac{q}{2}) +
\frac{G_2}{\kappa}\cos(k+\frac{q}{2})\bigr] }
{\sin(\frac{q}{2})\bigl[ \sin^2(\frac{\alpha}{2})\cos^2(\frac{q}{2}) -
\cos^2(\frac{\alpha}{2})\sin^2(\frac{q}{2}) \bigr]} \nonumber\\
&&\times\psi^{*}(k) \psi(k+q),
\label{displU}
\end{eqnarray}
\begin{equation}
S(q)\ =\ \sum_{k} \frac{\frac{\chi_2}{4\kappa_b}
\cos(\frac{\alpha}{2}) - \sin(\frac{\alpha}{2})
\cos(\frac{q}{2})\bigl[\frac{\chi_1}{\kappa}\cos(\frac{q}{2}) +
\frac{G_2}{\kappa}\cos(k+\frac{q}{2})\bigr] }
{ \sin^2(\frac{\alpha}{2})\cos^2(\frac{q}{2}) -
\cos^2(\frac{\alpha}{2})\sin^2(\frac{q}{2}) } \psi^{*}(k) \psi(k+q) .
\label{displS}
\end{equation}
Here $\psi(k)$ is given by \Ref{varphi2} in which $\psi(x)$, in turn,
is given by \Ref{dn-sol}.

\begin{figure}[htbp]
\unitlength1cm \hfil
\begin{picture}(8,12)
 \epsfxsize=8cm \put(-4,0){\epsffile{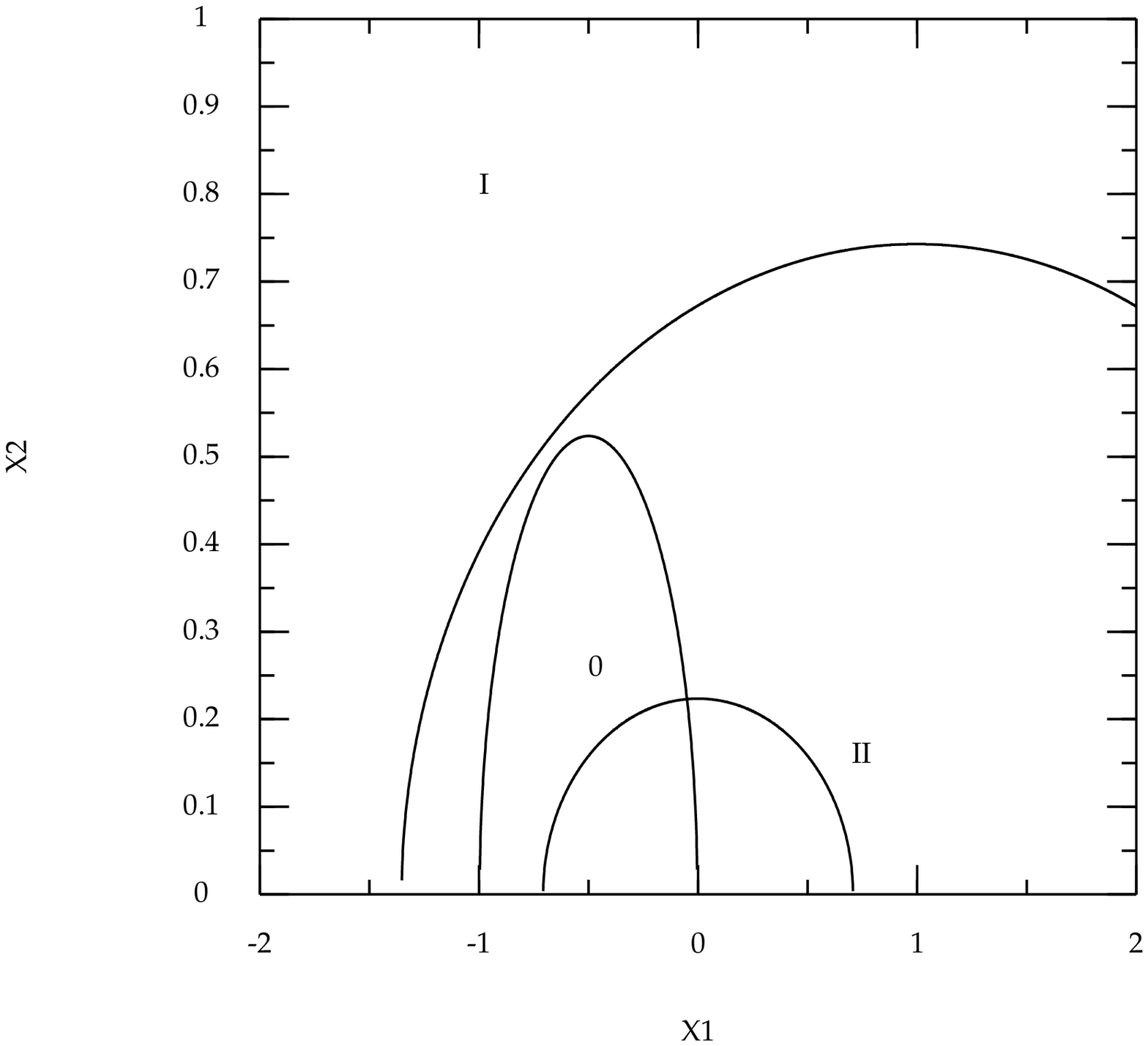}}
 \epsfxsize=8cm \put(4,0){\epsffile{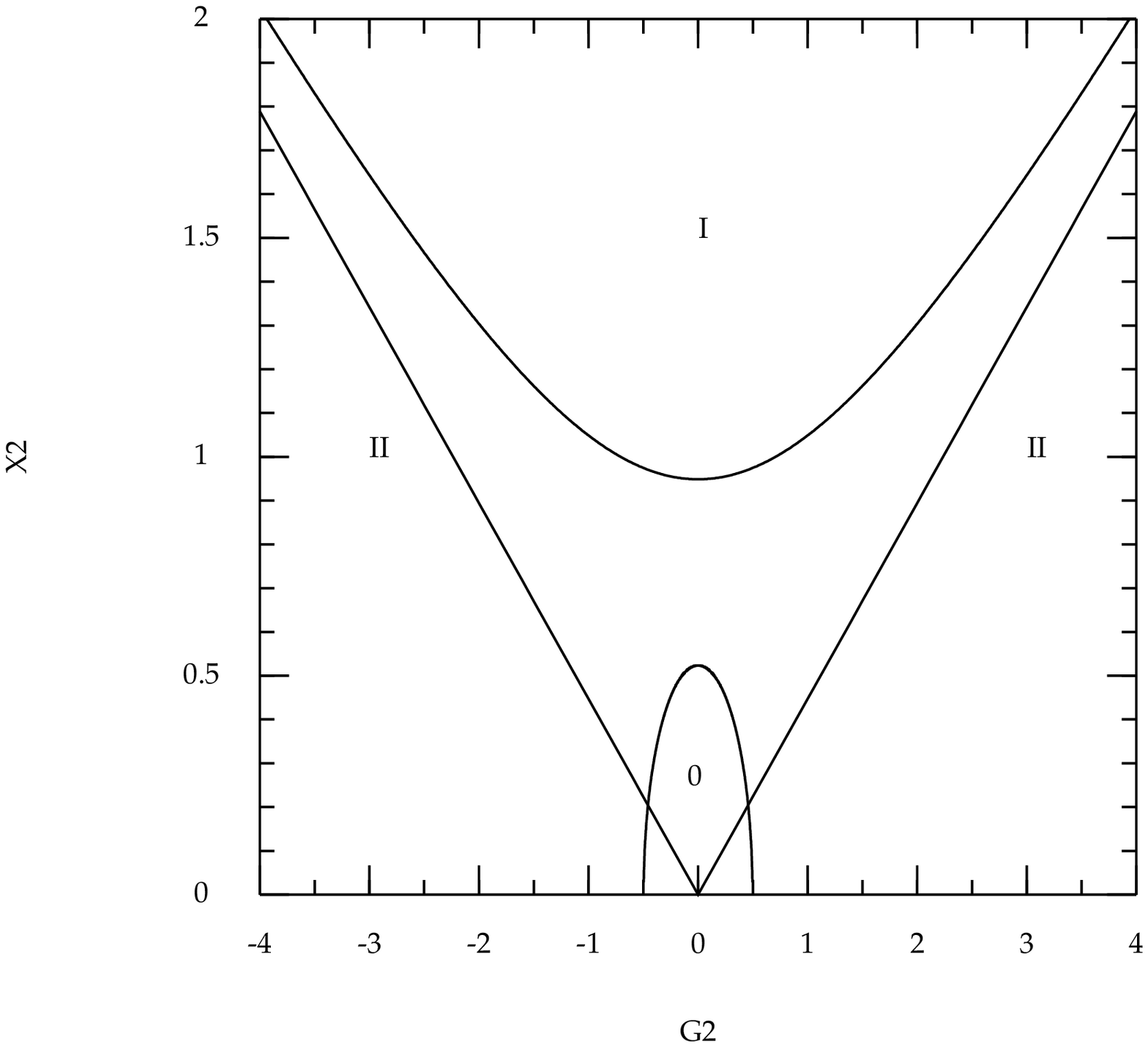}}
\put(1.5,0){a}
\put(7.5,0){b}
\end{picture}
\caption{\label{ApproxSol}
Regions of solutions: delocalised (0), strongly localised (I) and localised on 
two adjacent sites (II) for $G_2=0.5$ (a) and $\chi_1=0.5$ (b) }
\end{figure}

In Fig. \ref{ApproxSol} we present the curve of 
$g=g_{cr}=\pi^2/20$ for $N=20$, 
(\ref{g}), derived in the continuum approximation, where $G$ is given by
(\ref{approx}) with $b=0.5$. The curves on Fig. \ref{ApproxSol} are the two 
circles of radius $0.5$, but which look like vertical ellipses 
given the different scales of $x$ and $y$ in our plots, 
enclosing the region marked as '$0$'.
All the solutions inside these circles
(marked as $0$ in Fig \ref{ApproxSol}) are delocalised. 
Fig. \ref{ApproxSol}.a corresponds
to the case $G_2=0.5$ and  so can be compared to Fig. \ref{FIGplotG20.5} 
obtained numerically, while Fig \ref{ApproxSol}.b 
corresponds to the case $\chi_1=0$ - to be compared with Fig. \ref{FIGplotG2X2}

\section{The Limit of Strong Localisation}

As we have seen from the discussion above, the increase of the coupling 
constant
reduces the region of localisation of the quasiparticle. Thus, at a
large enough value of the coupling the quasiparticle can be localised 
mostly on one, two, to three
lattice sites. In such a case the conditions of the applicability of
 the continuum approximation are not satisfied.
To  consider the limit of strong localization, we use Eqs. (\ref{Gexpl})
-(\ref{F2-F1}) to rewrite Eq.(\ref{NLEqpsik}) as
\begin{eqnarray}
({\cal E}(k) &-& E + \frac{1}{N}\frac{{\chi_2}^2}{\kappa_b})\,\psi(k)\, +
\frac{1}{N}\frac{{\chi_2}^2}{\kappa_b}[\rho(\alpha)\psi(k-\alpha) +
\rho^*(\alpha)\psi(k+\alpha)]\nonumber\\
&-& \frac{1}{N} \sum_{q,k_1}G(k,k_1,q)
\psi^{*}(k_1) \psi(k_1+q) \psi(k-q)\,=\,0.
\label{NLEqpsik3}
\end{eqnarray}
Here, assuming  a  large enough circle with $N \gg 1$,  we have neglected
the term $\sim \sin(\alpha /4)$ in (\ref{F2-F1}). Moreover, we
 have taken into account the
normalization condition (\ref{norm}) and introduced the notation
\begin{equation}
\rho (q) = \sum_{k} \psi^{*}(k) \psi(k+q) .
\label{rho}
\end{equation}

The function $\psi(k) = (1/\sqrt{N})\sum_{n} \exp{(-ikn)}\phi _{n} $ is 
periodic in $k$, $\psi(k+2\pi) = \psi(k)$, and so it is representable as a
Fourier series
\begin{equation}
 \psi(k) = \frac{1}{\sqrt{N}} \bigl( c_0 + \sum_{m=1}^{\infty} c_m \cos (mk)
+ i \sum_{m=1}^{\infty} s_m \sin (mk) \bigr).
\label{psiFs}
\end{equation}
For $\rho (q)$ we then have
\begin{equation}
\rho (q) = c_0^2 + \frac{1}{2} \sum_{m=1}^{\infty}(c_m^2 + s_m^2) \cos(mq)
+ i \sum_{m=1}^{\infty} c_m s_m 
sin (mq) = \rho_r (q) + i\rho_i (q).
\label{rhoFs}
\end{equation}
The coeficients of the expansion satisfy the normalisation condition:
\begin{equation}
\sum_{k} |\psi (k)|^2 = \rho (0) = c_0^2 +
\frac{1}{2} \sum_{m=1}^{\infty}(c_m^2 + s_m^2) = 1.
\label{norm-cs}
\end{equation}

Next we put (\ref{psiFs}) into (\ref{NLEqpsik3}) and get the Fourier series
\begin{equation}
\{F(0)\,+\,\sum_{m}F(m)\cos (mk)\,+\,\sum_{m}\varphi(m)\sin (mk)\}\,=\,0,
\label{ff}
\end{equation}
thus showing that $F(m)=\varphi(n)=0$. These conditions give us the equations
for $c_m$ and $s_n$. We have
\begin{equation}
\Lambda_0c_0\,-\,J_+(0)c_1\,-\,T_-(0)s_1\,=\,0,
\end{equation}
\begin{equation}
\Lambda_1s_1\,-\,\frac{1}{2}J_+(2)s_2\,-\,N_1c_1\,-\,G_c(0)c_0\,-\,\frac{1}{2}
T_-(2)c_2\,=\,0,
\end{equation}
\begin{equation}
\Lambda_1c_1\,-\,J_c(0)c_0\,-\,\frac{1}{2}c_2\,-\,N_1s_1\,-\,\frac{1}{2}
T_-(2)s_2\,=\,0,
\label{cs}
\end{equation}
\begin{eqnarray}
\Lambda_mc_m\,&-&\,\frac{1}{2}J_-(m-1)c_{m-1}\,-\,\frac{1}{2}J_+(m+1)c_{m+1}
\,-\,N_ms_m\,\nonumber\\
&-&\,\frac{1}{2}T_+(m-1)s_{m-1}\,-\,\frac{1}{2}T_-(m+1)s_{m+1}\,=\,0,
\end{eqnarray}
\begin{eqnarray}
\Lambda_ms_m\,&-&\,\frac{1}{2}J_-(m-1)s_{m-1}\,-\,\frac{1}{2}J_+(m+1)s_{m+1}
\,-\,N_mc_m\,\nonumber\\
&-&\,\frac{1}{2}T_+(m-1)c_{m-1}\,-\,\frac{1}{2}T_-(m+1)c_{m+1}\,=\,0.
\end{eqnarray}
The functions $\Lambda_k$, $J_{\pm}(i)$, $T_{\pm}(j)$, $N_m$, $J_c(0)$ and
$G_c(0)$ are functions of $s_m$ and $c_n$ and so we see that we have nonlinear 
equations for $s_k$ and $c_i$.

Note that when we have a fully delocalised system all $c_k$ are the same; the 
same is true 
of $s_k$. As the coupling constant increases, $s_n$  become different from
 each other;
in fact, as $n$ increases, they become exponentially small (as a function of 
$n$). The same is true also for $c_n$. 
Thus, to get an approximate solution, we can take into account only a finite 
number of $s_i$ and $c_i$.

Our system of equations (\ref{cs}) always allows us to put $s_n=0$ and 
$c_n\ne0$.
If we now assume that $c_2 \ll c_1<c_0$, we can consider the equations 
involving only $c_0$ 
and $c_1$. In this case we have 
\begin{eqnarray}
&&\{\Lambda-(\frac{\chi_2^2}{\kappa_b}+\frac{2\chi_1^2}{\kappa})c_0^2-
\frac{\chi_1^2}{2\kappa}
c_1^2-\frac{8\chi_1G_2}{\pi \kappa}c_0c_1\}c_0\nonumber\\
&-&\{J+\frac{G_2^2}{\kappa}c_0c_1
 +\frac{4\chi_1G_2}{\pi\kappa}(c_0^2+\frac{1}{6})\}c_1\,=\,0,
\end{eqnarray}
\begin{eqnarray}
&& \{\Lambda-(\frac{\chi_2^2}{\kappa_b}
+\frac{2\chi_1^2}{\kappa})\frac{c_1^2}{4}-\frac{\chi_1^2}{\kappa}
c_0^2-\frac{8\chi_1G_2}{3\pi \kappa}c_0c_1\}c_1-\nonumber\\
&-&\{2J+\frac{2G_2^2}{\kappa}c_0c_1
+\frac{8\chi_1G_2}{\pi\kappa}(c_0^2+\frac{1}{6})\}c_0\,=\,0,
\label{c1s1}
\end{eqnarray}
where 
$\Lambda={\cal {E}}_0- E$.

When $c_1<c_0$ then  $c_0$ and $c_1$ are given by
\begin{equation}
c_1^2\,=\,\frac{(1+g_3)(D-g_1+g_2)}{D+g_3(2D+g_1-g_2)},
\label{c1sq}
\end{equation}
\begin{equation}
c_0^2\,=\,\frac{(1+3g_3)(D+g_1-g_2)}{2[D+g_3(2D+g_1-g_2)]},
\label{c0sq}
\end{equation}
where
$g_1=\frac{\chi_2^2}{\kappa_bJ}+\frac{2\chi_1^2}{\kappa J}$, 
$g_2=\frac{\chi_1^2}{\kappa J}+\frac{2G_2^2}{\kappa J}$, 
$g_3=4\frac{\chi_1 G_2}{\pi\kappa J}$. Here
$D$ is given by
 \begin{equation}
\label{D}
D=\sqrt{(g_1-g_2)^2+8(1+g_3)(1+3g_3)}.
\end{equation}
The quasiparticle wavefunction is now given by
\begin{equation}
\psi_n\,=\,c_0\delta_{n,n_0}\,+\,\frac{1}{2}c_1(\delta_{n,n_0+1}
+\delta_{n,n_0-1}),
\label{psi}
\end{equation}
and describes the localisation of the quasiparticle on at most 3 sites .

The energy of this state is given by $E={\cal {E}}_0-\frac{1}{2}\{g_1+g_2+D\}.$
Of course, our expression is valid when $c_1<c_0.$

In Fig. \ref{ApproxSol} we present the plots of the curve corresponding to 
$\frac{c_1^2}{c_0^2}=0.2$ 
obtained from (\ref{c1sq}), (\ref{c0sq}) and (\ref{D}).
In Fig. \ref{ApproxSol}.a we present the case $G_2=0.5$ as a function of 
$\chi_1$ and $\chi_2$ while Fig. \ref{ApproxSol}.b corresponds to
the case $\chi_1=0$ as a function of $\chi_2$ and 
$G_2$. Thus, our plots shows the curves which separate the regions of
strongly localised solutions from the region where the quasiparticle is 
localised on a few adjacent lattice sites. The curves are the ones that 
delimit 
the region marked as $I$ in the figure and which corresponds to the strongly 
localised solutions. When we compare Fig. \ref{ApproxSol}.a and 
Fig. \ref{ApproxSol}.b with Fig. \ref{FIGplotG20.5} and 
Fig. \ref{FIGplotG2X2}, 
respectively, we note a very good agreement between our numerical results
and our analytical results.

Finally, we consider the case where the quasiparticle is restricted to two 
neighbouring sites. In this case
we consider equations for $c_0$, $c_1$ and $s_1$ assuming that all other 
$s_i$ and $c_i$
are negligeably small. Thus, $c_2=O(\epsilon^2)$ and so is $s_2$.

In this case the equations are somewhat complicated, so here we present only 
their solutions.
They are given by:
\begin{equation}
c_1\,=\,c_0\,+\,O(\epsilon^2),\qquad s_1\,=\,c_0\,-\,O(\epsilon^2).
\end{equation}
The condition of normalisation gives us $c_0\sim \frac{1}{\sqrt{2}}.$
The quasiparticle wavefunction is then given by
\begin{equation}
\psi_n\,=\,\left(\frac{1}{\sqrt{2}}
         -O(\epsilon^2)\right)(\delta_{n,n_0}+\delta_{n,n_0+1}).
\end{equation}
The energy expression is now given by
\begin{equation}
E\,=\,{\cal {E}}_0 - J\,-\,\frac{1}{2}(g_1+g_2)\,+\,2g_3.
\end{equation}

Our numerical calculations, see Fig. 6c and d, have shown that
at the same values of the parameters we can have two different solutions, one 
localised essentially on two sites 
and the other localised over a larger number of sites. 
These two solutions have different energies.

The necessary condition for the existence of solutions localised
on two adjacent lattice sites is that $g_1-g_2$ satisfies the relation 
\begin{equation}
g_2-g_1\,=\,\frac{2G_2^2}{\kappa J}\,-
\,(\frac{\chi_2^2}{\kappa_b J}+\frac{\chi_1^2}{\kappa J})\,>0.
\label{eqdbl}
\end{equation}
The physically relevant solution is, of course, given by the state 
which has the lowest energy. The two site solution has the lowest
energy when $G_2$ is significantly larger than all other parameters.

In Fig. \ref{ApproxSol} we present the curve $g_2-g_1=0$ corresponding
to the limit of a solution localised on two adjacent sites. 
Fig. \ref{ApproxSol}.a presents the case of $G_2=0.5$ as a function of 
$\chi_1$ and $\chi_2$ and the curve is the ellipse, appearing as a circle 
on our scale, near the symbol 'II'.
The solutions localised on two sites exist just outside 
the ellipsoidal region marked as $II$. Note also that in the region contained
in the intersection between region $0$ and region $II$, the solutions are
delocalised. All this confirms the numerical results displayed in
Fig. \ref{FIGplotG20.5}.

Fig. \ref{ApproxSol}.b presents the case $\chi_1=0$ as a function of $\chi_2$ 
and $G_2$. The curves are the two straight lines and
the solutions localised on two sites exist in the region 
marked as $II$, just below the two lines. This is also in very good 
agreement with Fig. \ref{FIGplotG2X2}.

Notice also that we do not have any theoretical prediction for the transition 
from solutions localised on two sites  to those localised on three sites. 

For the case $G_2=0$, Eq. (\ref{eqdbl}) has no solution and so there are no
solutions localised on two neighbouring sites in that case, confirming our 
numerical results.

We should point out that we have also computed the theoretical predictions
for the different types of solutions corresponding to the case $G_2=0$ and 
$G_2=1$. When we have compared these predictions to the numerical results
presented in  Fig. \ref{FIGplotG20} and 
Fig. \ref{FIGplotG21} we have also found a very good agreement.

\section{Conclusions}

We have shown that one can successfully describe the electron-phonon
interaction of a circular chain while taking into account the transversal
and longitudinal displacement
of the ring of monomers. To achieve this, we have introduced a term in the
phonon Hamiltonian which describes the bending of the chain. This extra
displacement can, in turn, be coupled to the quasiparticle and this coupling
favours a self-trapping of quasiparticles and the formation of  
localised soliton-like structures.

The transversal displacement plays an important role in the formation of 
such states. The interaction between the quasiparticle and the 
phonon field leads not only to the localisation of the quasiparticle but 
also to a deformation of the circle. Depending on the coupling between 
the two fields, the circle is pinched at the position of the soliton 
or on the either side of it.

We have studied our model by solving its classical equations numerically. 
Our numerical studies have shown a great richness of solutions; hence we have
also tried to solve our equations analytically. To do this we have had
to perform  several approximations.  We have found that, at least
in the regions of validities of these approximations, our analytical
results are in good agreement with our numerical results.

Our results show that, depending of the choice of the parameters, our model
possesses
several classes of solutions. Fully delocalised solutions exist for all
values of the parameters. They do not correspond to the ground state 
when other solutions exist but they are the only solutions when the parameters
$\chi_1+G_2$ and $\chi_2$ are  small. On the other hand, strongly 
localised solutions exist when the coupling parameters are large. For these 
solutions, the quasiparticle is localised mostly on one lattice site, but 
the circle is completely distorted, {\it i.e.} the strong localisation applies 
only to the quasiparticle, not the displacement fields.

We also found some solutions that are localised on two adjacent sites and  
have a different symmetry.

Our study has shown that the electron-phonon interaction on a circular chain 
leads to the formation of bound states and to the deformation of the chain.
Our results can be used to the description of various systems mentioned 
in the introduction
and as a `warm-up' stage to the description of various 3-dimensional systems 
such as {\it e.g.} nanotubes.

\section{Acknowledgement}
This work has been supported by a Royal Society travel grant.
We would like to thank Yu. Gaididei for his interest and helpful discussions.

{}

\end{document}